\documentclass[%
 aip,
 amsmath,amssymb,
 reprint,%
]{revtex4-1}
\pdfoutput=1
\usepackage{hyperref}
\usepackage{epsfig}
\usepackage{amsmath}
\usepackage{graphicx}
\usepackage{wrapfig}
\usepackage{dcolumn}
\usepackage{bm}
\usepackage{amssymb}
\usepackage{titlesec}
\usepackage{mathtools}
\usepackage{relsize}
\usepackage{cleveref}
\usepackage{calligra}
\usepackage{lipsum}

\usepackage[usenames,dvipsnames]{color}

\begin{document}

\title{Large and small fluctuations in oscillator networks from heterogeneous and correlated noise}
\author{Jason Hindes$^{1}$, Ira B. Schwartz}
\affiliation{U.S. Naval Research Laboratory, Washington, DC 20375, USA}
\author{Melvyn Tyloo}
\affiliation{Theoretical Division and Center for Nonlinear Studies (CNLS), Los Alamos National Laboratory, Los Alamos, NM 87545, USA}
\date{\today}
\begin{abstract}
Oscillatory networks subjected to noise are broadly used to model physical and technological systems. Due to their nonlinear coupling, such networks typically have multiple stable and unstable states that a network might visit due to noise. In this manuscript, we focus on the assessment of fluctuations 
resulting from heterogeneous and correlated noise inputs on Kuramoto model networks. We evaluate the typical, small fluctuations near synchronized states and connect the network variance to the overlap between stable modes of synchronization and the input noise covariance. Going beyond small to large fluctuations, we introduce the \textit{indicator mode approximation}, that projects the dynamics onto a single amplitude dimension.
Such an approximation allows for estimating rates of fluctuations to saddle instabilities, resulting in phase slips between connected oscillators. Statistics for both regimes are quantified in terms of effective noise amplitudes that are compared and contrasted for several noise models. Bridging the gap between small and large fluctuations, we show that a larger network variance does not necessarily lead to higher rates of large fluctuations.
\end{abstract} 
\maketitle

\begin{quotation}
    Noise unavoidably affects complex networked systems, inducing small excursions around stable collective states, and eventually, leading to basin escape and transitions from one collective state to another. In most analyses, noise inputs are assumed to be homogeneous in intensities and uncorrelated among the network components. However, such assumptions are typically not satisfied in many real systems and therefore, represent only a first approximation. Here, we go beyond this approximation and investigate the effect of both heterogeneous and correlated noise on the two regimes of small and large fluctuations in Kuramoto oscillators networks.
\end{quotation}

\section{\label{sec:Intro}INTRODUCTION}
Coupled oscillators play a crucial role in the modelling of various physical and engineered systems ranging from power grids to neuronal dynamics~\cite{strogatz2001exploring,pikovsky2002synchronization,acebron2005kuramoto,RevModPhys.94.015005}. The combination of their coupling and internal dynamics gives rise to collective behaviors such as synchronization, in which the coupled units exhibit coherent dynamics that is often oscillatory~\cite{ARENAS200893,rodrigues2016kuramoto,dorflerPNAS2013}. Typically, the interplay of the coupling network topology and the inherent nonlinearity of interactions produces multiple stable and unstable states that exist for the same set of internal parameters\cite{acebron2005kuramoto,RevModPhys.94.015005,rodrigues2016kuramoto,10.1063/1.4978697,10.1063/1.4927292}. In particular, for phase oscillator networks with sufficiently large coupling, states in which all oscillators rotate collectively with the same frequency (called synchronized states for short) can be represented by network fixed points, each with their own basin of attraction\cite{wiley2006size,menck2013basin,DORFLER20141539,10.1063/1.4978697,10.1063/1.4986156}. Due to external perturbations or noise an oscillator network will explore its initial basin of attraction. If the noise is small, the excursions will remain in the vicinity of the initial synchronized state for relatively long time scales compared to local relaxation times. However, eventually noise and the nonlinear dynamics will organize in such a way as to drive a network to points of instability, resulting in possible transitions from one basin to another\cite{kramers1940brownian,dykman1990large,hindes2018rare,hindes2019network}. In either case, it is important to assess the induced fluctuations and possible transitions, as they can harmfully impact the dynamics of network systems~\cite{hines2009cascading}.

Noisy oscillatory networks have been investigated from various angles, focusing for example, on the relation between fluctuations and topology~\cite{ronellenfitsch2018optimal,tyloo2018robustness,10.1063/5.0122898}, noise propagation~\cite{zhang2020vulnerability,tyloo2022layered}, noise-induced synchronization~\cite{Mar22}, and network escape and large fluctuations resulting from Gaussian and non-Gaussian noise~\cite{deville2012transitions,PhysRevE.95.060203,hindes2018rare,hindes2019network,Ritmeester_2022,tyloo2023finite}. However, most works have assumed homogeneous and/or uncorrelated noise. Yet, heterogeneity and correlation are important realistic ingredients in physical and spatially embedded networks. Indeed, since the physical parameters in models of oscillator networks have heterogeneity, e.g., in the natural frequencies and the coupling topology, we should not expect homogeneity in the noise as well, particularly if the noise derives from these parameters. For instance in power-grid models, fluctuations typically arise from the input power at each node, which has a mean proportional to an oscillator's natural frequency\cite{bergen1981structure}. If the mean is different on the nodes, it is reasonable to expect the temporal variation around the mean to be as well. Also, correlation should be expected from the geographical positions and coupling of a network's components, as well as from unmodelled interactions between oscillators. Recent works have explored small fluctuations emerging from noise with time and space correlations, and addressed how to devise networks with e.g., spatial embedding that optimally cancels noise and correlation patterns that optimally improve network coherence\cite{ronellenfitsch2018optimal,PhysRevLett.128.098301}. Here we aim to further consider heterogeneous noise intensities and go beyond to include large fluctuations. In particular, we analyze the response of phase oscillator networks in both regimes of small fluctuations (SFs) around synchronized states and large fluctuations (LFs), eventually leading to saddle instabilities and phase slips between connected oscillators. We also clarify the relation between SFs and LFs, namely do SFs give any information about eventual transitions? We show that, surprisingly, larger network variance in the SF regime does not always translate into shorter escape times in the LF regime. 

The dynamical system we consider is a network of Kuramoto oscillators subjected to noise, whose time-evolution is governed by a set of differential equations~\cite{10.1007/BFb0013365},
\begin{align}
\label{eq:Kuramoto}
&\dot{\theta_{i}}=\omega_{i}+K\sum_{j}A_{ij}\sin\!\big(\theta_{j}-\theta_{i}\big)+\sum_{m}G_{im}\xi_{m}(t), 
\end{align}
where $i\!\in\!\{1,2...,N\}$\cite{acebron2005kuramoto,ARENAS200893}. In Eq.(\ref{eq:Kuramoto}) $\omega_{i}$ is the natural frequency of the $i$-th oscillator\footnote{We note that throughout this work in simulations, we draw the natural frequencies for each node independently from a uniform distribution over the interval $[-0.25,0.25]$ for simplicity.}, and the coupling between oscillators is given by an adjacency matrix $A_{ij}$ multiplied by a coupling strength $K$\cite{acebron2005kuramoto,ARENAS200893,DORFLER20141539}. The noise sources $\{\xi_{1},\xi_{2},...,\xi_{M}\}$ are i.i.d. white Gaussian processes with $\left<\xi_{m}(t)\xi_{n}(t')\right>\!=\!\sigma^{2}\delta_{mn}\delta(t-t')$ and a feed-in matrix $G$. Given $G$, the covariance for the input noise is $GG^{T}$, which can encode both heterogeneous intensities and correlation patterns. Our goal is to understand how different intensity distributions and correlations in the noise implicit in $G$ affect the dynamics of Eq.(\ref{eq:Kuramoto}), and in particular, affect the SFs and LFs from synchronized states.

This manuscript proceeds as follows: in Sec. \ref{sec:SF}, we perform a linear analysis of Eq.(\ref{eq:Kuramoto}) around the synchronized state, which allows us to determine the noise-induced variance for general $G$. In subsections \ref{sec:Het}-\ref{sec:Mode} we further analyze the SF variance within the context of heterogeneous and uncorrelated noise on the nodes, correlated edge noise, and collective-mode noise. For each noise model, we extract analytical insight by finding the variances in limiting cases of weak stability, large coupling, and special model networks. In Sec.\ref{sec:LF}, we estimate the rate of large fluctuations to saddle instabilities by constructing a single-mode approximation and then solve for the mode amplitude’s large deviations. In subsection \ref{sec:Compare}, we compare the effective amplitudes of LFs and SFs for the noise models introduced in subsections \ref{sec:Het}-\ref{sec:Mode} and show that, typically, they are only moderately correlated. In addition, we indicate when LF rates are maximized for the different noise models. Sec.\ref{sec:Conclusion} offers a summary and discussion of future research directions. 

\section{\label{sec:SF}SMALL FLUCTUATIONS}
To begin, let us explore how noise drives the network dynamics to produce SFs. If the coupling is sufficiently large, Eq.(\ref{eq:Kuramoto}) admits a stable state of phase-locked synchronization, which is a fixed-point, $\boldsymbol{\theta}^{*}$, satisfying $\omega_{i}\!-\!\bar{\omega}\!+\!K\sum_{j}A_{ij}\sin(\theta_j^*\!-\!\theta_i^*)\!=\!0 \;\forall i$\cite{DORFLER20141539}. Without loss of generality, we assume the average frequency is zero, $\bar{\omega}\!=\!\sum_{i}\omega_{i}/N\!=\!0$, where we can always redefine the natural frequencies relative to the average. If the noise intensities are small enough, the response of the network is well approximated by the first-order Taylor expansion of Eq.(\ref{eq:Kuramoto}) around the synchronized state, $\boldsymbol{\theta}(t)\!=\!\boldsymbol{\theta}^{*} +\boldsymbol{\delta\theta}(t)$:
\begin{equation}\label{eq1}
    \dot{\delta\theta}_i = -K\,\sum_j A_{ij}\cos(\theta_i^*-\theta_j^*)(\delta\theta_i - \delta\theta_j) + \sum_m G_{im}\,\xi_m(t)\,.
\end{equation}
Defining the Laplacian matrix,
\begin{eqnarray}\label{eq6}
   \tilde{\mathbb{L}}_{ij} = 
   \begin{cases}
           -\,A_{ij}\cos(\theta_i^*-\theta_j^*) \,, i\neq j\,,\\
            \,\sum_{k}A_{ik}\cos(\theta_i^*-\theta_k^{*}) \,, i=j\,,
   \end{cases}
\end{eqnarray}
and denoting its eigenvectors $\bm u_{\alpha}$ with $u_{1,i}\!=\!1/\sqrt{N}$\, and eigenvalues $\lambda_1\!=\!0\!<\!\lambda_2\!<\!...\!<\!\lambda_N$\,, one can expand the response as $\delta\theta_i(t) \!=\! \sum_\alpha \!c_\alpha(t)u_{\alpha,i}$\,. One should notice that deviations of the phases along ${\bm u_1}$ do not modify the dynamics. Thus, we only focus on deviations orthogonal to ${\bm u_1}$. The general solution of Eq.(\ref{eq1}) is given by 
\begin{eqnarray}
    \label{eq:TimeDep}
    \delta\theta_i(t)=\sum_\alpha e^{-\lambda_\alpha\,\!K\,t}\!\!\!\int_0^t \!e^{\lambda_\alpha\,\!K\,t'}\!\sum_{k,m} G_{mk}\,\xi_k\, u_{\alpha,m}\,{\rm d}t'\, u_{\alpha,i}.
\end{eqnarray}

The latter expression is useful to calculate the moments of  SFs\cite{tyloo2018robustness}, e.g. $\langle \delta\theta_i^2(t\rightarrow \infty) \rangle$ which converges to a finite value after a transient of order $\lambda_{2}^{-1}$\,. As we are interested in the interplay between network topology and noise heterogeneity and correlation, and not in the local dynamics, we focus on the {\it total} network variance, or the sum of the node variances $\langle{ \boldsymbol{\delta \theta}^{T}\boldsymbol{\delta \theta}\rangle}$. We can write the total network variance in a useful form as proportional to an effective noise amplitude for small fluctuations, $q$:
\begin{equation}
\label{eq:SFs}
\langle{ \boldsymbol{\delta \theta}^{T}\boldsymbol{\delta \theta}\rangle}=\dfrac{\sigma^{2}}{2}q, \;\;\text{where}\;\; q=\frac{1}{K}\sum_{\alpha}\frac{{\bm u_{\alpha}^{T}}GG^T{\bm u_{\alpha}}}{\lambda_\alpha}.
\end{equation}


The SF noise amplitude, $q$, gives us a simple multiplicative factor for how the feed-in noise is effectively amplified by the network to produce its noise-induced variance. Equation (\ref{eq:SFs}) shows us that SFs around the synchronized state depend crucially on the covariance matrix of the noise, $GG^{T}$, with larger variances produced when the principal modes of the covariance align with the slowest modes of the synchronized state. Next, we explore Eq.(\ref{eq:SFs}) further with specific noise models.


\subsection{\label{sec:Het}Heterogeneous and uncorrelated noise}
\begin{figure}[h]
\center{\includegraphics[scale=0.202]{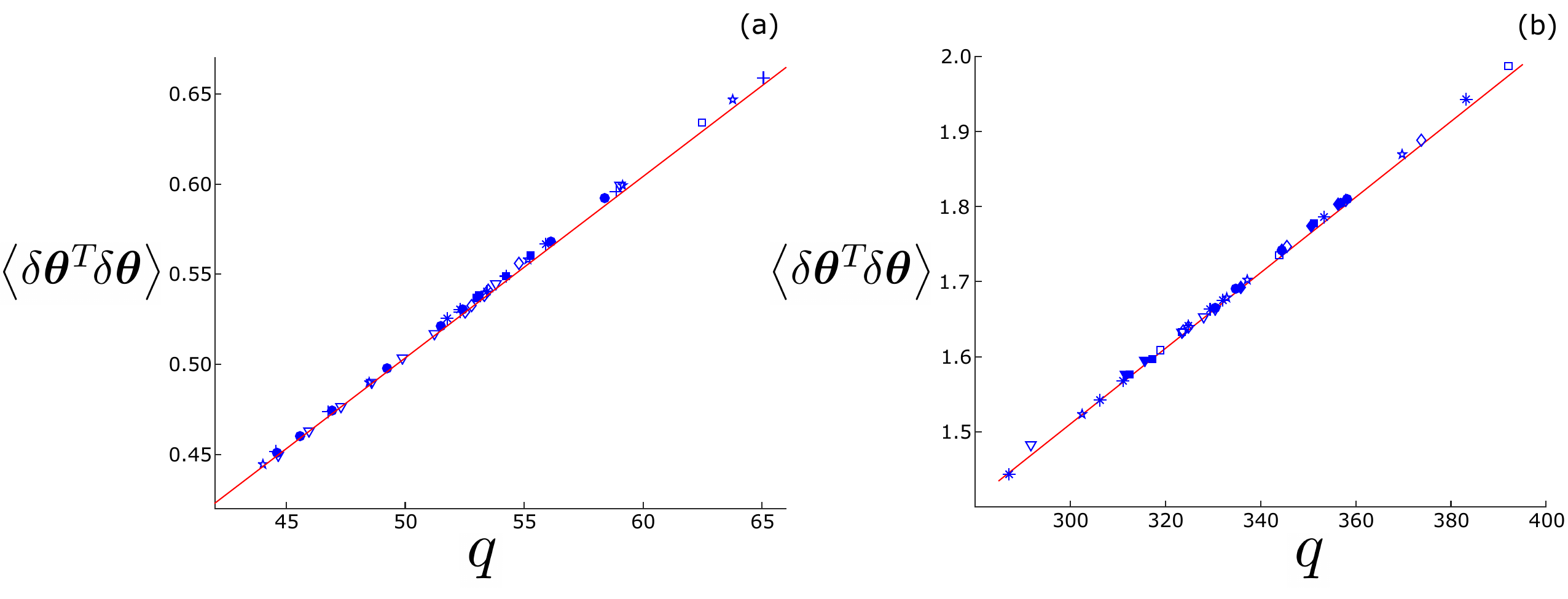}}
\caption{Network variance versus small fluctuation amplitude from simulations of Eq.(\ref{eq:Kuramoto}) with heterogeneous and uncorrelated noise on the nodes. 
Each point represents a random realization of the noise intensities which are i.i.d. (a) IEEE-30 test bus network with $K\!=\!1.6K_{SN}$ and $\sigma^{2}\!=\!0.02$; (b) UK power grid with $K\!=\!1.4K_{SN}$ and $\sigma^{2}\!=\!0.01$, where $K_{SN}$ is the coupling at which the synchronized state emerges. Predictions from Eq.(\ref{eq:SFs}) are drawn with red lines.}
\label{fig1}
\end{figure}

Let us begin with heterogeneous and uncorrelated noise on the nodes, for which $[GG^T]_{ij}\!=\! \delta_{ij}\,g_{i}^2$. First, we test the general Eq.(\ref{eq:SFs}) in Fig.\ref{fig1}, where we plot the total network variance observed in stochastic simulations of Eq.(\ref{eq:Kuramoto}) versus the effective SF amplitude $q$ for two networks: (a) the IEEE-30 test grid, and (b) the UK power grid~\cite{simonsen2008transient}. For the noise intensities, we use a simple model with independent and uniform distributions, $g_{i}^{2}\!\sim\!1\!+\!h *U_{i}(-0.5,0.5)$, where $h$ is a heterogeneity parameter, and $U_{i}(-0.5,0.5)$ is the uniform distribution for the $i$th oscillator over the interval $[-0.5,0.5]$. Different symbols in Fig.\ref{fig1} correspond to different random samples from the $U_{i}$'s. For a given sample, the heterogeneity parameter $h$ is also varied while keeping $g_{i}^2\!>\!0$ $\forall i$. In Fig.\ref{fig1} we see that the SF predictions from Eq.(\ref{eq:SFs}), shown with a red line, are in good agreement with simulations and in particular confirm the linearity of the total network variance with $q$ for both networks.  

We can gain further analytical insight on the effect of heterogeneity with uncorrelated noise by considering noise intensities that are randomly determined and i.i.d., as in the example Fig.\ref{fig1}. In particular, let us assume that that the first and second moments of the noise-intensity distribution are respectively $\overline{g_{i}^2}\!=\!\mu$ and $\overline{g_{i}^4}\!=\!s^{2}\!+\!\mu^{2}$. In this case, one has for the average and variance of Eq.(\ref{eq:SFs}):
\begin{align}
\label{eq:moments}
\begin{split}
      \mu_{q} &= \frac{ \mu}{K}\sum_\alpha \frac{1}{\lambda_\alpha}\, \;\;\; \text{and}\\
     \sigma_{q}^{2} &= \frac{s^{2}}{K^2}\sum_{\alpha,\beta} \frac{\sum_ku_{\alpha,k}^2u_{\beta,k}^2}{\lambda_\alpha\lambda_\beta}.
     \end{split}
\end{align}

Firstly, note that in Eq.(\ref{eq:moments}) the expectation of $q$ is equivalent to the standard homogeneous noise case where the intensities are identical to their means. In general, the result implies larger expected SFs for weakly connected networks, which only depend on the sum of the inverse stability eigenvalues. 
We can get a sense for the importance of noise heterogeneity by considering $\sigma_{q}^{2}/\mu_{q}^{2}$. First, near the emergence of the synchronized state at the saddle-node bifurcation, $K\!\gtrsim\!K_{SN}$, Eq.(\ref{eq:moments}) is dominated by the slowest mode $\alpha\!=\!2$ (called the Fiedler mode) with $\lambda_{2}\!\approx\!0$\cite{hindes2018rare,hindes2019network}. In this case, $\sigma_{q}^{2}/\mu_{q}^{2}\!\approx\!(s^{2}/\mu^{2})\sum_{i}u_{2,i}^{4}$. It is known that for typical saddle-node bifurcations, the Fiedler mode effectively splits the oscillator network into two subgraphs that lose synchrony at bifurcation\cite{hindes2018rare,hindes2019network}. If we denote the sizes of the two subgraphs $W$ and $N\!-\!W$, we have $u_{2,i}\!\approx\!-\sqrt{(N\!-\!W)/WN}$ if $i$ is in the smaller subgraph and $u_{2,i}\!\approx\!\sqrt{W/(N-W)N}$ otherwise; note that the results are exact for so called, single-cut saddle node bifurcations\cite{hindes2019network,Ritmeester_2022}. If $W\!\ll\!N$, then $\sigma_{q}^{2}/\mu_{q}^{2}\!\approx\!(s^{2}/\mu^{2})/W$. For example, for the UK grid in Fig.(\ref{fig1})(b), $\sigma_{q}^{2}/\mu_{q}^{2}\!=\!(s^{2}/\mu^{2})/7.52$, which is close to the expected value with $W\!=\!7$. We conclude that when the number of weakly connected oscillators $(W)$ is small, noise heterogeneity effects can be significant, even for large networks.

On the other hand if the coupling is large $K\!\gg\!K_{SN}$, $\cos(\theta_i^{*}\!-\!\theta_j^{*})\!\simeq\!1$. In this case, the weighted network Laplacian, $\tilde{\mathbb{L}}$, becomes identical to the unweighted network Laplacian, $\mathbb{L}$\cite{tyloo2019key}. To understand Eq.(\ref{eq:moments}) in the large coupling limit further, we consider special networks with known eigenstructure\cite{mieghem_2010,newman2018networks}. For instance, for complete, star, and circle graphs we have $\sum_{\alpha}\!\lambda_{\alpha}^{-1}\!\simeq\!  1, N, N^2/12$ and $\sum_{\alpha,\beta} \frac{\sum_ku_{\alpha,k}^2u_{\beta,k}^2}{\lambda_\alpha\lambda_\beta}\!\simeq\!  1/N, N, \;\text{and}\; N^3/144$\, respectively for large $N$. From this we observe that dense networks tend to be insensitive to noise heterogeneity as their variance in $q$ goes to zero with $N$\,. Intuitively, such networks have short and multiple paths among the nodes which makes them efficient in averaging-out local disorder. In contrast, networks with few paths among nodes, including networks with central hubs and networks with large average path lengths, have variances in $q$ that tend to grow with $N$, and are thus poor at averaging-out heterogeneity.
Nevertheless, despite their differences, all example networks have $\sigma_{q}^{2}/\mu_{q}^{2}\!\simeq\!(s^{2}/\mu^{2})/N$, implying that we can expect noise heterogeneity effects to be small in relative terms in the limit of large networks and coupling.

\subsection{\label{sec:Edge}Edge-correlated noise}
Next, we consider noise that is correlated/anti-correlated through edges of the coupling network. We start with the case where noise is {\it perfectly anti-correlated} for each edge. Anti-correlation between pairs of connected nodes can emerge, for instance, if there is additive noise in the coupling $K$, since the sinusoidal interaction in Eq.(\ref{eq:Kuramoto}) is an odd function. Perfectly anti-correlated edge noise can be achieved by choosing $G\!=\!B$ where $B$ is the signed incidence matrix such that if the $m$th edge connects nodes $i$ and $j$\,, i.e. $A_{ij}\!>\!0$\,, then $B_{im}\!=\!1$\,, $B_{jm}\!=\!-1$ and $B_{km}\!=\!0$ otherwise (where one has to choose an orientation for each edge). With this definition, the noise covariance matrix reads as $GG^T\!=\!BB^T\!=\!\mathbb{L}$\cite{newman2018networks}. 
As a consequence, the SF amplitude is
\begin{align}\label{eq:AntiCorrEdgeGen}
\begin{split}
     q^{(-)} = \frac{1}{K}\sum_{\alpha}\frac{{\bm u_{\alpha}^{T}}\mathbb{L}{\bm u_{\alpha}}}{\lambda_\alpha}\,,
\end{split}
\end{align}
where $(-)$ signifies anti-correlated edge noise.

For further analytical perspective on Eq.(\ref{eq:AntiCorrEdgeGen}), we once more consider the large coupling limit where $\tilde{\mathbb{L}}\!\simeq\! \mathbb{L}$\, whereby Eq.(\ref{eq:AntiCorrEdgeGen}) reduces to  
\begin{align}\label{eq:varanti2}
\begin{split}
     q^{(-)} \simeq \frac{(N-1)}{K}\,.
\end{split}
\end{align}
Interestingly, for anti-correlated noise on every edge, the total network variance is predicted to depend on the network size but not the topology when the coupling is large.  

\begin{figure}[h]
\center{\includegraphics[scale=0.195]{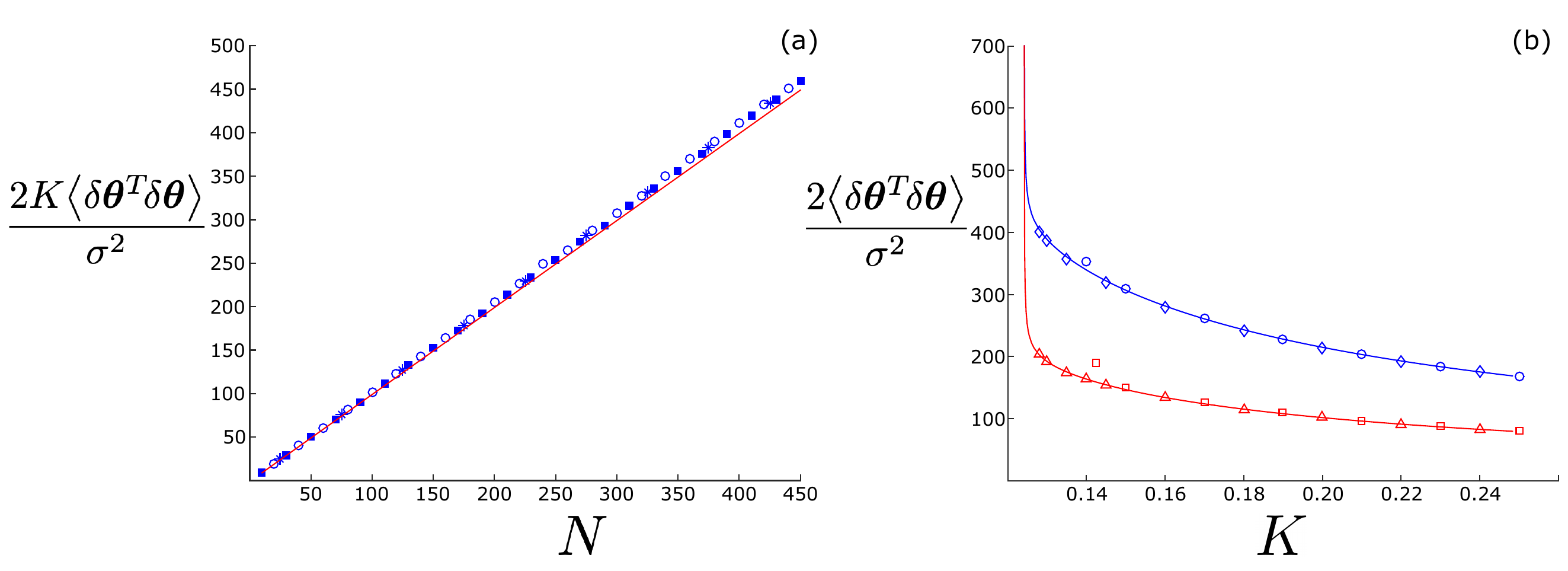}}
\caption{Network variance from correlated edge noise. (a) Normalized variance versus network size for Watts-Strogatz (circles), Barabasi-Albert (squares), and Erd\H{o}s-Renyi (asterisks) networks for anti-correlated edge noise with $K\!=\!0.6\approx 2.5K_{SN}$ and $\sigma^{2}\!=\!0.02$. Predictions from Eq.(\ref{eq:varanti2}) are drawn with a red line. (b) Normalized variance versus coupling for a 20-node Watts-Strogatz network. Blue circles ($\sigma^{2}\!=\!10^{-3}$) and diamonds ($\sigma^{2}\!=\!10^{-4}$) correspond to simulations of Eq.(\ref{eq:Kuramoto}) with correlated edge noise while red squares ($\sigma^{2}\!=\!10^{-3}$) and triangles ($\sigma^{2}\!=\!10^{-4}$) correspond to anti-correlated edge noise. Predictions from Eq.(\ref{eq:SFs}) are drawn with solid curves. 
}
\label{fig2}
\end{figure}

Example predictions for anti-correlated edge noise are demonstrated in Fig.\ref{fig2} (a). In particular, we fix the coupling strength to $K\!\approx\!2.5K_{SN}$ and vary the sizes of networks subjected to anti-correlated noise. Three different classes of networks are plotted: Watts-Strogatz with $\left< k\right>\!=\!4$ and $10\%$ random connections (circles), Barabasi-Albert with ``new" nodes added with degree $m\!=\!4$ (squares), and Erd\H{o}s-Renyi with $\left< k\right>\!=\!4$ (asterisks)\cite{newman2018networks}. Each random network model produces very different topologies. 
Yet, as predicted by Eq.(\ref{eq:varanti2}), all network variances are proportional to the number of nodes only. 

The opposite case of perfectly correlated edge noise is slightly more involved, and can arise in the presence of small noisy phase frustration, since the derivative of the sine interaction is symmetric, e.g., $\sin(\theta_{j}^{*}\!-\!\theta_{i}^{*}\!+\!w(t))\!\approx\!\sin(\theta_{j}^{*}\!-\!\theta_{i}^{*})\!+\!w(t)\!*\cos(\theta_{j}^{*}\!-\!\theta_{i}^{*})$\cite{sakaguchi1988mutual}. For correlated edge noise the feed-in matrix is given by $G\!=\!\tilde{B}$ where $\tilde{B}$ is the unsigned incidence matrix such that if the $m$th edge connects nodes $i$ and $j$\,, i.e. $A_{ij}\!>\!0$\,, then $\tilde{B}_{im}\!=\!\tilde{B}_{jm}\!=\!1$\,, and $\tilde{B}_{km}\!=\!0$ otherwise. In this case 
\begin{align}\label{eq:varanti}
\begin{split}
     q^{(+)}&= q^{(-)}+\frac{2}{K}\!\sum_{\alpha}\frac{{\bm u_{\alpha}^{T}}A{\bm u_{\alpha}}}{\lambda_\alpha},
\end{split}
\end{align}
where $(+)$ signifies correlated edge noise.

Again, assuming that the relative angles at the fixed point are small, Eq.(\ref{eq:varanti}) implies that the difference between correlated and anti-correlated edge noise depends on how the principal modes of the network adjacency matrix overlap with the slowest modes of the Laplacian. For an analytical perspective, we once more make use of known spectra for model networks\cite{mieghem_2010,newman2018networks}. For instance for complete, star, and circle graphs, we can compute $[q^{(+)}-q^{(-)}]\big/K\!\simeq\! -2, -4/N,\;\text{and}\; N^{2}/3$, respectively for large $N$. From these special cases, we make the general observations that for networks with small average path lengths, e.g., dense networks and networks with central hubs, correlated and anti-correlated edge noise (or mixtures of both) tend to produce nearly identical noise-induced variances with $q\!\simeq\!N/K$. On the other hand, we expect variances to be much larger for correlated versus anti-correlated noise, if the underlying network is weakly connected, and in particular, has a large diameter. 

Predictions for correlated and anti-correlated edge noise are displayed in Fig.\ref{fig2} (b). We plot the normalized variance for correlated (top) and anti-correlated (bottom) edge noise for two noise intensities as a function of the coupling strength for a Watts-Strogatz network\cite{watts1998collective}. We can see that the variance quickly increases as we approach the bifurcation $K_{SN}\!\approx\!0.125$. 
On the other hand, as the coupling increases both approach the expected behavior, $q\!\rightarrow\!\text{{\it constant}}/K$ for $K\!\gg\!K_{SN}$. Note that the correlated series is above the anti-correlated, which is consistent with our expectations from Eq.(\ref{eq:varanti}), since the underlying network is sparse and without high-degree hubs.

\subsection{\label{sec:Mode}Collective-mode noise}
Finally, before moving on to large fluctuations, we consider noise that is correlated along the modes of the synchronized state, i.e $G_{im}\!=\! G_{0,m}u_{m,i}$ for $m=2,...,M$\,, where $M\!\ll\!N$. We call this case collective mode noise. From Eq.(\ref{eq:SFs}), the SF network amplitude reads
\begin{align}\label{eq:var_modes}
\begin{split}
     q = \frac{1}{K}\sum_{m}\frac{G_{0,m}^2}{\lambda_m}\,.
\end{split}
\end{align}
Because of its simple connection to the network modes, Eq.(\ref{eq:var_modes}) is easy to interpret. For instance, if all the modes have the same intensity $G_{0,m}\!=\!G_{0}$, the $l$th mode contributes a fraction $\lambda_{l}^{-1}/\sum_{m}\lambda_{m}^{-1}$ of the variance. 
More generally, if the noise intensity is held constant, $I\!=\!\sum_{m}G_{0,m}^2$, we can see that the network variance is  {\it maximized} when $G_{0,2}^{2}\!=\!I$. 
Namely, the largest SFs occur when we ``pump" all the noise into the
Fiedler mode.


\section{\label{sec:LF}LARGE FLUCTUATIONS}
Next, we turn to the problem of large fluctuations (LFs) from synchronized states, and in particular, fluctuations that drive oscillator networks
to unstable saddle points\cite{kramers1940brownian,freidlin2012random,deville2012transitions,hindes2018rare,doi:10.1137/17M1142028}. Upon reaching a saddle, the usual pattern for Kuramoto model networks is for some number of connected oscillators 
to develop full phase slips with respect to each other before returning to a synchronized state\cite{deville2012transitions,hindes2018rare}. For instance, in the simple case of tree networks, 
the two subgraphs that would be disconnected by the removal of a given edge (the edge associated with the saddle instability) slip\cite{manik2014supply,doi:10.1137/120899728,hindes2019network}.

\textcolor{black}{Generally for such LFs, the statistics are not Gaussian, which is the case within the SF approximation. In particular, in the small-noise limit the probability distributions for LFs are effectively exponential with exponents that are not simple quadratic forms (as in the SF approximation), but more complex functions reflecting the full nonlinear dynamics.} The probability exponents for LFs are called ``actions", since they are describebale in terms of classical mechanics~\cite{kramers1940brownian,freidlin2012random,dykman1990large,doi:10.1137/17M1142028,Assaf_2017}. In general, analytical solutions for actions are unknown, and one must resort to solving a two-point boundary value problem numerically in an augmented Hamiltonian system~{\cite{lindley2013iterative}}. However, this approach is numerically intensive, and therefore slow, and often fails to provide analytical insights for high-dimensional networked systems.

Here, we develop an analytical approximation technique for estimating large fluctuation rates for Eq.(\ref{eq:Kuramoto}) based on mode projection. Our approach is to look for solutions of the dynamics Eq.(\ref{eq:Kuramoto}) where all of the oscillators move together according to $\theta_{i}(t)\!=\!\theta_{i}^{*}+a(t)\Delta\theta_{i}$ $\forall i$, 
where $\Delta\theta_{i}\!\equiv\! \theta_{i}^{s}-\theta_{i}^{*}$, and where $\boldsymbol{\theta}^{s}$ is the saddle fixed point of Eq.(\ref{eq:Kuramoto}). Note that the saddle instability emerges, in general, with the stable synchronized state at the saddle-node bifurcation $K\!=\!K_{SN}$. Typical of mode projections, we substitute this ansatz into Eq.(\ref{eq:Kuramoto}), multiply the $i$th equation by $\Delta\theta_{i}$, sum over all $i$, and solve for $\dot{a}$. The result is the following stochastic differential equation for the projection amplitude $a(t)$:
\begin{align}
\label{eq:IMA}
&\dot{a}= F(a)+ \dfrac{\sum_{i}\Delta\theta_{i}\sum_{m}G_{im}\xi_{m}(t)}{\sum_{l}\Delta\theta_{l}^{2}},\;\;\;\; \text{where} \nonumber \\
&F(a)=\!\dfrac{\!\sum_{i}\!\Delta\theta_{i}\!\Big[\omega_{i} +\!K\!\sum_{j}\!A_{ij}\sin\!\big(\theta_{j}^{*}\!-\!\theta_{i}^{*}+\!a(\Delta\theta_{j}\!-\!\Delta\theta_{i})\!\big)\!\Big]}{\sum_{l}\Delta\theta_{l}^{2}}.
\end{align}
Note that despite its seeming complexity, Eq.(\ref{eq:IMA}) has the simple form of 
a one-dimensional, nonlinear system with additive noise.  

Before continuing our analysis, let us point out two benefits of the projection. First, we encode the exact fixed-points of the network-- the stable synchronized state 
and the saddle -- into the dynamics of $a$. Hence, Eq.(\ref{eq:IMA}) has conserved the exact boundary conditions for LFs but in a lower-dimensional 
theory. Incidentally, we call the approach the {\it indicator mode approximation} (IMA), because $a\!=\!0$ encodes the synchronized state, while $a\!=\!1$ ``indicates" that an LF has occurred.   
Second, as we approach the saddle-node bifurcation, $|\theta_{i}^{s}-\theta_{i}^{*}|\!\ll\!1$ $\forall i$, the IMA dynamics is expected to be increasingly accurate, since the oscillators can be shown 
to follow an effective one-dimensional dynamics where $\Delta\boldsymbol{\theta}$ is proportional to the Fiedler mode of the synchronized state~\cite{manik2014supply,hindes2019network}. 

Before uncovering several other useful features of the IMA, we point out that the goal in front of us is to construct the augmented Hamiltonian system for describing LFs of the mode projection amplitude $a$. To do so, we need to construct the Fokker-Planck equation for the probability flux of $a$, which requires understanding the noise term in Eq.(\ref{eq:IMA}). To make progress, note that we have a linear summation of i.i.d Gaussian random variables. Hence, the noise term on the right hand side of Eq.(\ref{eq:IMA}) will have a Gaussian distribution 
with some mean and variance. Let us define the effective noise $\eta(t)\equiv\sum_{i}\Delta\theta_{i}\sum_{m}G_{im}\xi_{m}(t)\big/\sum_{l}\Delta\theta_{l}^{2}$. It is straightforward to see that the mean of $\eta$ is zero, since $\left<\xi_{m}\right>\!=\!0$ $\forall m$. 
To calculate the variance of $\eta$ we use two well-known properties of random variables: 1) the variance of the sum of random variables is the sum of the variances. 2) The variance of a constant multiplied by a random variable is the variance of the random variable multiplied by the constant squared {\cite{ross2014first}}. As a consequence,
\begin{align}
\label{eq:IMAvariance}
&\left<\eta(t)\eta(t')\right>=\sigma^{2}\sum_{m}Q_{m}^{2}\;\delta(t-t'), \;\;\; \text{where} \nonumber \\
&Q_{m}=\dfrac{\sum_{i}G_{im}\Delta\theta_{i}}{\sum_{l}\Delta\theta_{l}^{2}}.      
\end{align}

Given Eqs. (\ref{eq:IMA}) and (\ref{eq:IMAvariance}), we can now write the Fokker-Planck equation for the probability flux of $a$,  
\begin{align}
\label{eq:IMFP}  
\dfrac{\partial P}{\partial t} = -\dfrac{\partial}{\partial a}\big[F(a)P\big] +\dfrac{1}{2}\dfrac{\partial^{2}}{\partial a^{2}}\big[\sigma^{2}\!\sum_{m}Q_{m}^{2}\;P\big].
\end{align}
Next, we look for solutions of Eq.(\ref{eq:IMFP}) in the WKB form, $P\!\sim\!\exp[-2S(a,t)/\sigma^{2}]$, 
which is the expected form for describing large deviations of $a$ in the tail of the probability distribution\cite{freidlin2012random,billings2010switching,dykman2010poisson,doi:10.1137/17M1142028,Assaf_2017}. We note that the WKB approximation is generally valid in the limit $\sigma^{2}\!\ll\!1$. 
If we plug the WKB ansatz into Eq.(\ref{eq:IMFP}), and take the limit $\sigma^{2}\!\rightarrow\!0$, we convert the Fokker-Planck equation into a Hamilton-Jacobi equation, $\partial{S}/\partial{t}\!+\!H\!=\!0$, at $\mathcal{O}(1/\sigma^{2})$ with 
Hamiltonian: 
\begin{align}
\label{eq:IMH}  
H= F(a)\dfrac{\partial S}{\partial a} +\sum_{m}Q_{m}^{2}\Big(\dfrac{\partial S}{\partial a}\Big)^{2}.  
\end{align}
Equation (\ref{eq:IMH}) is the standard form for the LF Hamiltonian of a one-dimensional nonlinear process with additive Gaussian white noise\cite{freidlin2012random,doi:10.1137/17M1142028}. Since we are looking for fluctuations from a stable synchronized state, we want to solve for the stationary distribution where $S(a,t)$ (called the {\it action}) has no explicit time dependence, $\partial{S}/\partial{t}=0$. In this case $H\!=\!0$, and therefore 
\begin{align}
\label{eq:IMM}  
\dfrac{\partial S}{\partial a}=\dfrac{d S}{d a}= -\dfrac{F(a)}{\sum_{m}Q_{m}^{2}}.  
\end{align}
Finally, integrating Eq.(\ref{eq:IMM}) from the synchronized state $a\!=\!0$ to the saddle $a\!=\!1$, we derive the IMA action, which is an approximation for the probability-exponent for LFs to a saddle (denoted $S$, henceforth, for brevity): 
\begin{align}
\label{eq:IMS}  
S\;\;=\;&\dfrac{\sum_{l}\Delta\theta_{l}^{2}}{\sum_{m,i}(G_{mi}\Delta\theta_{i})^{2}}\;S_{DB}, \;\;\;\; \text{where} \nonumber\\
S_{\text{DB}}\;\;=\;&\sum_{i}\Delta\theta_{i}\Big[\!-\!\omega_{i} +K\!\!\!\!\!\!\!\sum_{\substack{j, \\ \Delta\theta_{i}\neq\Delta\theta_{j}}}\!\!\!\!\!\dfrac{A_{ij}}{\Delta\theta_{j}\!-\!\Delta\theta_{i}}\cdot\big(\cos(\theta_{j}^{s}\!-\!\theta_{i}^{s}) \nonumber \\
&-\cos(\theta_{j}^{*}\!-\!\theta_{i}^{*})\!\Big]. 
\end{align}

We have written the action in a suggestive form in Eq.(\ref{eq:IMS}). Namely, within the IMA the action takes the form of the {\it detailed balance limit} action, re-scaled by the inverse of an effective noise amplitude, $Q$:  
\begin{align}
\label{eq:Q}  
Q=\dfrac{\Delta\boldsymbol{\theta}^{T}GG^{T}\Delta\boldsymbol{\theta}}{\Delta\boldsymbol{\theta}^{T}\Delta\boldsymbol{\theta}}.
\end{align}
Recall that for Kuramoto model networks with i.i.d. additive noise on the nodes (where $G$ is an $N\text{x}N$ identity matrix), the LF paths are identical to the time-reversed paths of the noise-free system, and have actions given by $S_{\text{DB}}$\textcolor{black}{\cite{hindes2018rare,PhysRevE.95.060203}}. Hence, the IMA produces the exact result for homogeneous noise, which is true, not just near bifurcation as we expect, but in general. We point out that this is interesting, since it is easy to check that the time-reversed paths of the deterministic system do not satisfy the constrained, mode assumptions of the IMA. Nevertheless, the IMA produces the correct action. \textcolor{black}{We further point out for comparison that, even in the case of homogeneous noise intensities on the nodes for which the IMA gives the exact probability exponent to reach a saddle in the small-noise limit, the SF (Gaussian) approximation is $\sum_{\alpha}\!\frac{1}{2}K\lambda_{\alpha}(\Delta\boldsymbol{\theta}^{T} \bm{u}_{\alpha})^{2}$. One can check that this SF result has a significant fractional error of $\gtrsim50\%$.}    

As previously mentioned, the probabilities for LFs described by large-deviations theory generally take the form $P\!\sim\!\exp[-S/D]$, where $D$ is an effective noise amplitude for the considered process\textcolor{black}{\cite{freidlin2012random,doi:10.1137/17M1142028,billings2010switching}}. In our case, we have $D\!=\!\sigma^{2}Q/2$. Note that similar to the the SF network amplitude, $q$, all of the feed-in noise dependence for LFs is in $Q$. Namely, if we keep all parameters constant (the topology, the coupling, the natural frequencies, and the input noise intensities) while changing the noise feed-in matrix alone, changes in the probabilities for LFs are contained within changes in $Q$. In particular, as we increase $Q$, we increase the likelihood for LFs (within the IMA). 

\subsection{\label{sec:Compare}Noise model numerical comparisons}
Similar to Sec.\ref{sec:SF}, our approach for the remainder of this section is to understand and test the predictions of Eq.(\ref{eq:IMS}) using the noise models presented in Sec.\ref{sec:SF}, and then compare $Q$ and $q$ for each. In order to test our IMA with simulations of Eq.(\ref{eq:Kuramoto}), we use a standard observable for LFs: the average waiting time, $\left< T \right>$ \textcolor{black}{\cite{doi:10.1137/17M1142028,kramers1940brownian}}. In general, LFs in the small-noise limit are expected to be Poisson processes with exponentially distributed waiting times\textcolor{black}{\cite{doi:10.1137/17M1142028,kramers1940brownian}}. Moreover, the observation rate is expected to be proportional to the probability, or $\left< T \right>\!=\!B*\exp[-2S_{DB}/\sigma^{2}Q]$ where $B$ is a slowly-varying prefactor function compared to $\exp[-2S_{DB}/\sigma^{2}Q]$. Thus, for large $\left< T \right>$ we can predict $\ln\!\left< T \right>$ to within a constant,  
\begin{align}
\label{eq:Times}  
\ln\left< T \right>\approx \dfrac{2S_{DB}}{\sigma^{2}Q} + \text{constant}.  
\end{align} 
To compare to Eq.(\ref{eq:Times}), we measure the time it takes for a $2\pi$ phase-difference to appear between connected oscillators in simulations of Eq.(\ref{eq:Kuramoto}), average the times over $200$ stochastic realizations, and then repeat for different $G$ within the noise-model classes defined in Sec.\ref{sec:SF}. Then, we plot our results with an additive constant. 
 
\begin{figure}[h]
\center{\includegraphics[scale=0.209]{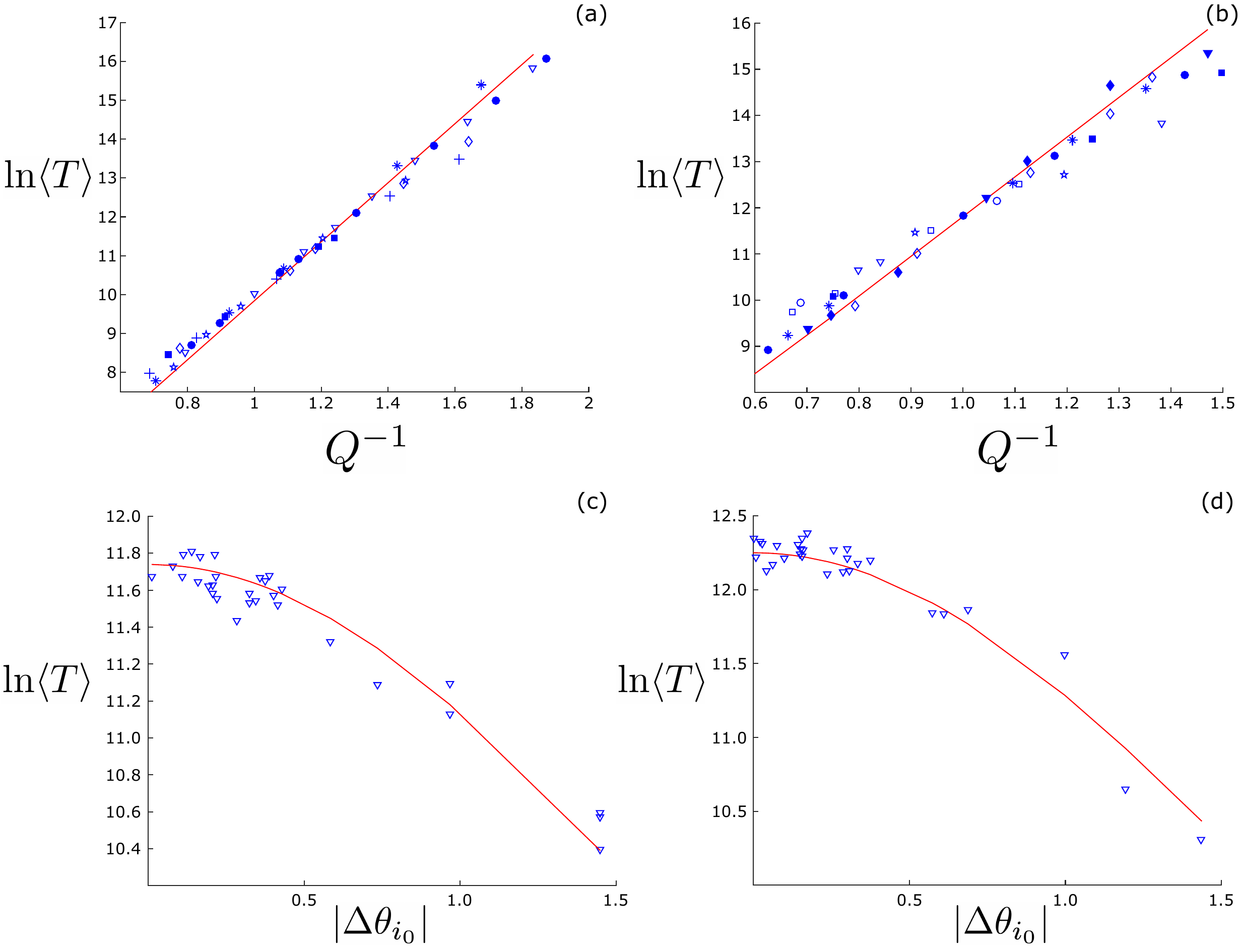}}
\caption{Natural logarithm of network slip times from uncorrelated and heterogeneous noise on the nodes. Slip times versus the inverse of the large fluctuation amplitude for: (a) the IEEE-30 test bus network with $\sigma^{2}\!=\!0.1$, and  (b) the UK power grid with $\sigma^{2}\!=\!0.05$. The noise distributions, coupling, and simulation plot-labels for (a-b) follow the same convention as for Fig.\ref{fig1}. Slip times versus the distance a targeted node, $i_{0}$, must travel to a saddle for: (c) the IEEE-30 test bus network with $K\!=\!1.6K_{SN}$ and $\sigma^{2}\!=\!0.08$, and (d) a 3-regular random network with $K\!=\!1.5K_{SN}$, $\sigma^{2}\!=\!0.07$, and $N\!=\!100$. For (c-d) the noise intensities on the nodes are all equal, except for a single targeted node that has doubled intensity. Simulation points are drawn with blue triangles. In all panels, predictions from Eq.(\ref{eq:Times}) are drawn with red lines.}
\label{fig3}
\end{figure}

We start with the case where noise is uncorrelated and heterogeneous on the nodes, $[GG^T]_{ij}\!=\! \delta_{ij}\,g_{i}^2$. In this model, Eq.(\ref{eq:Q}) becomes
\begin{align}
\label{eq:Qhet}
Q= \dfrac{\sum_{l}g_{l}^{2}\Delta\theta_{l}^{2}}{\sum_{i}\Delta\theta_{i}^{2}}. 
\end{align}
Following Fig.\ref{fig1}, we can plot simulated slip times using the same networks and random noise intensities on the nodes as Fig.\ref{fig1}; results are shown in Fig.\ref{fig3} (a-b) for the same two example networks. First, we notice that we quantitatively capture the dependence of the average slip times as we change the underlying noise distribution, according to Eqs.(\ref{eq:IMS}), (\ref{eq:Times}), and (\ref{eq:Qhet}). Second, we notice that the ordering of points in Fig.\ref{fig1} and Fig.\ref{fig3} (a-b) are not identical,  meaning that the fastest slip rates do not necessarily correspond to the largest network variances. A simple measure for the correspondence between variances and LF rates is the Pearson correlation coefficient, $R\!=\!\text{cov}(q,Q)\big/\text{var}(Q)\text{var}(q)$, where $\text{cov}(q,Q)$ is the covariance of $q$ and $Q$, and $\text{var}(q)$ and $\text{var}(Q)$ are the respective variances over the samples of $GG^{T}$. Though simple and linear, this choice is natural since $q$ and $Q$ are both amplitudes for the linear operator $GG^{T}$. For the data plotted in Fig.\ref{fig1}(a-b) and Fig.\ref{fig3}(a-b) we have for the two example networks \textcolor{black}{$R\!=\!0.83$} and \textcolor{black}{$R\!=\!0.84$}, respectively.
Therefore, in the case of random intensity distributions, we find a strong, but not perfect, correlation between the total network variance and the rate of LFs.  

But, what about non-random intensities? In particular, Eq.(\ref{eq:Qhet}) implies that the largest response for LFs comes from pumping noise into the nodes that must travel the farthest to the saddle. Namely, if we fix the noise intensity, $I\!=\!\sum_{i}g_{i}^{2}$, then we expect a maximum rate for LFs when 
$g_{l}^{2}\!=\!I$ where $l\!=\!\text{max}_{i}\{\Delta\theta_{i}^{2}\}$, i.e., not necessarily when the Fiedler mode is targeted. 
We can test the model of node-targeted noise by assuming that the noise-intensities on the nodes are identical except for a single node, denoted $i_{0}$, that has, e.g., double the intensity. This case is plotted for two example networks\footnote{For $k$-regular random networks all nodes have the same degree, $k$, but neighbors for each node are selected uniformly at random.} in Fig.\ref{fig3} (c-d). Though there is some spread around the simple IMA prediction, the expected behavior is demonstrated: a decrease in the average slip times with the saddle distance of the targeted node. As with Fig.\ref{fig3}(a-b), we can repeat the Pearson computation comparing noise-induced SFs and LFs. Interestingly, we find significantly lower degrees of correlation between the total network variance and the rate of LFs with \textcolor{black}{$R\!=\!0.41$} and \textcolor{black}{$R\!=\!0.26$} for (c) and (d), respectively.

Next, we consider edge correlated noise. Similar to heterogeneous noise-intensities on the nodes, we focus on $GG^{T}$ that tend to align with fluctuations to $\boldsymbol{\theta}^{s}$. As such, let us consider {\it mixtures} of perfectly correlated and anti-correlated edge noise. Namely, we assume that if the $m$th edge connects nodes $i$ and $j$, then $G_{mi}\!=\!1$ and $G_{mj}\!=\!r_{ij}$ where $r_{ij}\in\{-1,1\}$. In this case Eq.(\ref{eq:Q}) becomes
\begin{align}
\label{eq:Qedge}
Q=\dfrac{\sum_{i}\!k_{i}\Delta\theta_{i}^{2} +\sum_{i,j}\!A_{ij}r_{ij}\Delta\theta_{i}\Delta\theta_{j}}{\sum_{l}\!\Delta\theta_{l}^{2}},
\end{align}
where $k_{i}$ is the degree of the $i$th node. Given Eq.(\ref{eq:Qedge}), it is easy to see that, in order to increase the rates of LFs, we want the edge correlations to align with the product of the displacements to the saddle for connected nodes. In particular, in order to maximize the rates of LFs:
\begin{align}
\label{eq:Max}
r_{ij}^{\text{(max)}}=\text{sign}(\Delta\theta_{i}\Delta\theta_{j}), \;\; \forall\{i,j\}\;\; \text{with}\;\; A_{ij}=1. 
\end{align}

\begin{figure}{h}
\center{\includegraphics[scale=0.209]{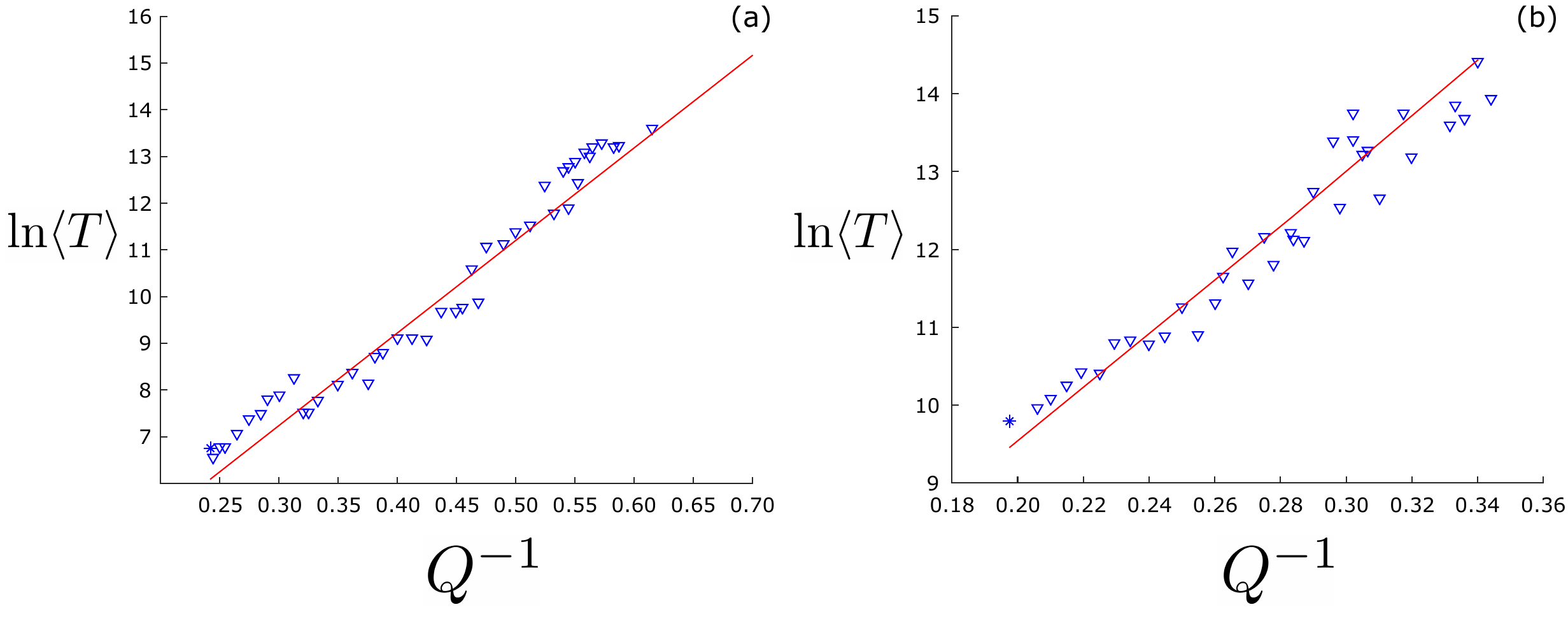}}
\caption{Natural logarithm of network slip times from random mixtures of correlated and anti-correlated edge noise versus the inverse of the large fluctuation amplitude for: (a) a 20-node Watts-Strogatz network with $K\!=\!1.5K_{SN}$ and $\sigma^{2}\!=\!0.02$, and (b) a 100-node 3-regular network with $K\!=\!1.5K_{SN}$ and $\sigma^{2}\!=\!0.02$. Simulation points are drawn with blue triangles while predictions from Eq.(\ref{eq:Times}) are drawn with red lines. The mixture of edge noise that is predicted to have the minimum slip time is drawn with an astericks for both (a) and (b).}
\label{fig4}
\end{figure}

Example slip times from correlated edge noise are plotted in Fig.\ref{fig4} for two networks, panels (a) and (b). For each point, we have independently and randomly assigned a correlation ($1$ or $-1$) to every edge in the network. Hence, different points are different random sample realizations of the mixture correlated edge-noise model. As with the heterogeneous noise examples, we can see that the combined 
Eqs.(\ref{eq:IMS}), (\ref{eq:Times}), and (\ref{eq:Qedge}) capture the quantitative behavior of average slip times as the noise correlations are varied. Moreover, the predicted minimum slip times (having the maximum slip rate), which are plotted with astericks in each panel, appear very near the bottom of the sampled slip times, in agreement with the IMA prediction Eq.(\ref{eq:Max}). In order to compare SFs and LFs in this edge-noise model, we can compute the Pearson correlations for the example networks in Fig.\ref{fig4} (a-b), which are $R\!=\!0.21$ and $R\!=\!0.24$, respectively. Therefore we find that, similar to targeted node noise, we have a relatively low degree of correlation between the noise-induced network variance and the rate of LFs when noise is correlated along the edges of a network.

\begin{figure}[h]
\center{\includegraphics[scale=0.209]{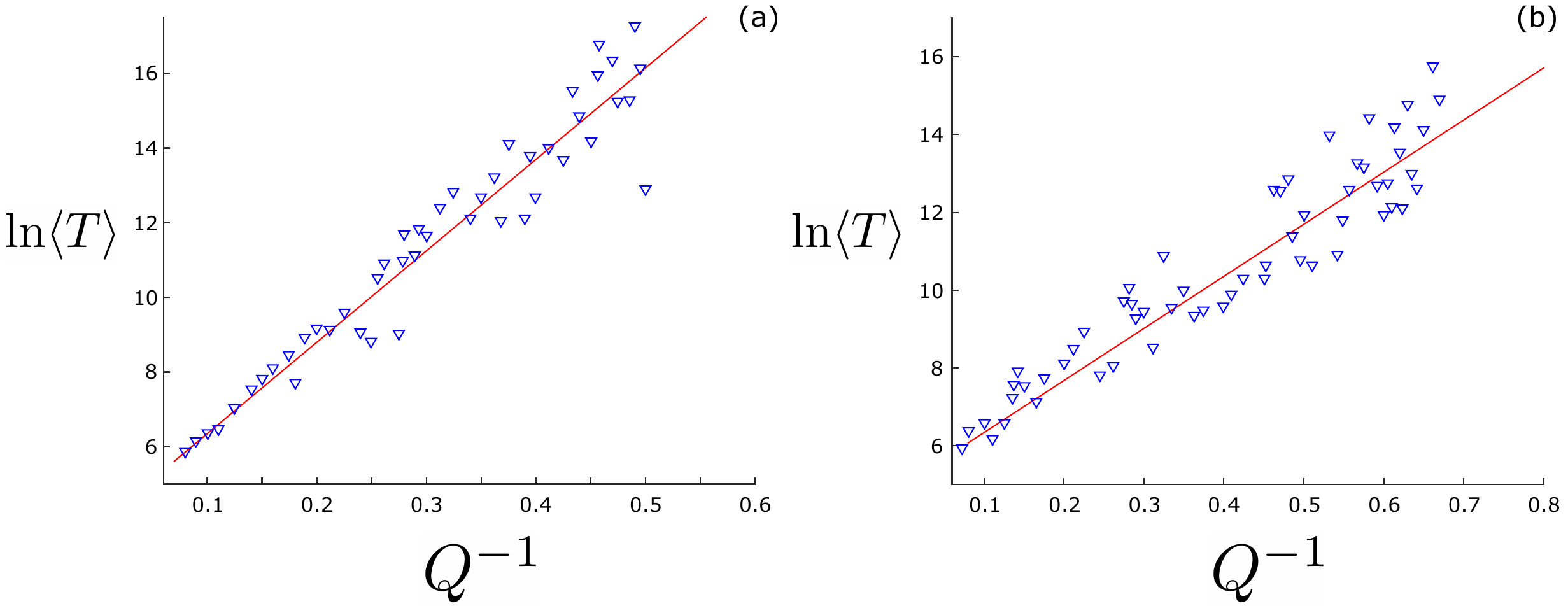}}
\caption{Natural logarithm of network slip times from collective mode noise versus the inverse of the large fluctuation amplitude for: (a) the IEEE-30 test bus network with $K\!=\!2K_{SN}$ and $\sigma^{2}\!=\!0.06$, and (b) the UK power grid with with $K\!=\!1.4K_{SN}$ and $\sigma^{2}\!=\!0.02$. Simulation points are drawn with blue triangles while predictions from Eq.(\ref{eq:Times}) are drawn with red lines. For each simulation point, the noise intensities for each mode are drawn randomly.} 
\label{fig5}
\end{figure}

The final example that we consider is collective-mode noise. In this model, the noise is introduced through a relatively small number of the network’s stable modes of the synchronized state, with $G_{im}\!=\! G_{0,m}u_{m,i}$ as in Sec.\ref{sec:Mode}. 
For comparisons to simulations we chose the five slowest modes of the IEEE network and the UK power grid, since they produce the largest variances from Eq.(\ref{eq:var_modes}). Examples are shown in Fig.\ref{fig5} (a-b). Each point represents the average slip times from random intensities on the modes, which are sampled from i.i.d uniform distributions, $G_{0,m}\!\sim\!U_{m}(-4,4)$. Here, we can see that the IMA predictions plotted in red capture the trend, though the spread around our estimate predictions are appreciable in this highly-correlated noise model. For the Pearson correlation between $q$ and $Q$, we find $R\!=\!0.63$ and $R\!=\!0.69$ for (a) and (b), respectively, for the two example networks. Such values indicate a moderate correlation between the total network variance and the rate of LFs for collective-mode noise. In fact, the level of correlation is similar to the model of uncorrelated and random noise intensities on the nodes, though not as high.

\section{\label{sec:Conclusion}CONCLUSION}
Noise is an inevitable component of the dynamics of complex networks,
which typically couple together many heterogeneous dynamical systems. Important examples are oscillator networks, which are used to model the synchronization of electric power networks, neuronal networks,  Josephson junction arrays, and many other physical and biological systems. Since the physical parameters that govern the dynamics of such systems are typically heterogeneous and spatially embedded, and often not all of the interactions between components can be fully modeled, we expect the noise on oscillator networks to be both heterogeneous and correlated. If the noise is small, a network has small fluctuations on short time scales around stable states, while on long times scales, noise can create fluctuations to unstable saddle states and result in basin escape. 

In this work, we addressed both large and small fluctuations from synchronized states in coupled phase oscillator networks, and showed that both can be quantified in terms of distinct effective noise amplitudes for each regime. In the small fluctuation regime, we found that noise heterogeneity tends to persist for weakly stable networks, networks with central hubs, and networks with large diameters. On the other hand, for noise that is correlated along edges of the network, the network variance for large networks tends to depend only on the number of nodes in the network as the coupling strength is increased, except for networks with large diameters and positive correlations. In the large fluctuation regime we found that noise maximizes the rate of large fluctuations when it targets nodes that must travel the farthest to saddles, and when noise is correlated along edges of a network such that the correlation between nodes is aligned with the product of their displacements to a saddle. Comparing the two regimes we showed that the total network variance and the rate of large fluctuations can be significantly correlated in the case of randomly heterogeneous noise, while only weakly correlated for targeted and edge-correlated noise. 

Future work will consider extensions of the indicator mode approximation for estimating large fluctuation rates, to include more network modes and increased numerical accuracy. In addition, interesting avenues for broader application of the work presented could entail, for example, analyzing large and small fluctuations of noisy oscillator networks with only partial synchronization, and in more complex network dynamical systems including: electric power systems, swarming networks, and swarmalator systems.

\section{\label{sec:Acknowledgements}ACKNOWLEDGEMENTS}
\noindent JH and IBS were supported by the U.S. Naval
Research Laboratory funding (N0001419WX00055) and the
Office of Naval Research funding (N0001419WX01166) and (N0001419WX01322). MT was supported by the Laboratory Directed Research and
Development program of Los Alamos National Laboratory under project number 20220797PRD2 and by U.S. DOE/OE as part of
the DOE Advanced Sensor and Data Analytics Program.

\bibliography{bibliography}

\begin{thebibliography}{48}%
\makeatletter
\providecommand \@ifxundefined [1]{%
 \@ifx{#1\undefined}
}%
\providecommand \@ifnum [1]{%
 \ifnum #1\expandafter \@firstoftwo
 \else \expandafter \@secondoftwo
 \fi
}%
\providecommand \@ifx [1]{%
 \ifx #1\expandafter \@firstoftwo
 \else \expandafter \@secondoftwo
 \fi
}%
\providecommand \natexlab [1]{#1}%
\providecommand \enquote  [1]{``#1''}%
\providecommand \bibnamefont  [1]{#1}%
\providecommand \bibfnamefont [1]{#1}%
\providecommand \citenamefont [1]{#1}%
\providecommand \href@noop [0]{\@secondoftwo}%
\providecommand \href [0]{\begingroup \@sanitize@url \@href}%
\providecommand \@href[1]{\@@startlink{#1}\@@href}%
\providecommand \@@href[1]{\endgroup#1\@@endlink}%
\providecommand \@sanitize@url [0]{\catcode `\\12\catcode `\$12\catcode
  `\&12\catcode `\#12\catcode `\^12\catcode `\_12\catcode `\%12\relax}%
\providecommand \@@startlink[1]{}%
\providecommand \@@endlink[0]{}%
\providecommand \url  [0]{\begingroup\@sanitize@url \@url }%
\providecommand \@url [1]{\endgroup\@href {#1}{\urlprefix }}%
\providecommand \urlprefix  [0]{URL }%
\providecommand \Eprint [0]{\href }%
\providecommand \doibase [0]{http://dx.doi.org/}%
\providecommand \selectlanguage [0]{\@gobble}%
\providecommand \bibinfo  [0]{\@secondoftwo}%
\providecommand \bibfield  [0]{\@secondoftwo}%
\providecommand \translation [1]{[#1]}%
\providecommand \BibitemOpen [0]{}%
\providecommand \bibitemStop [0]{}%
\providecommand \bibitemNoStop [0]{.\EOS\space}%
\providecommand \EOS [0]{\spacefactor3000\relax}%
\providecommand \BibitemShut  [1]{\csname bibitem#1\endcsname}%
\let\auto@bib@innerbib\@empty
\bibitem [{\citenamefont {Strogatz}(2001)}]{strogatz2001exploring}%
  \BibitemOpen
  \bibfield  {author} {\bibinfo {author} {\bibfnamefont {S.~H.}\ \bibnamefont
  {Strogatz}},\ }\href@noop {} {\bibfield  {journal} {\bibinfo  {journal}
  {Nature}\ }\textbf {\bibinfo {volume} {410}},\ \bibinfo {pages} {268}
  (\bibinfo {year} {2001})}\BibitemShut {NoStop}%
\bibitem [{\citenamefont {Pikovsky}\ \emph {et~al.}(2002)\citenamefont
  {Pikovsky}, \citenamefont {Rosenblum},\ and\ \citenamefont
  {Kurths}}]{pikovsky2002synchronization}%
  \BibitemOpen
  \bibfield  {author} {\bibinfo {author} {\bibfnamefont {A.}~\bibnamefont
  {Pikovsky}}, \bibinfo {author} {\bibfnamefont {M.}~\bibnamefont {Rosenblum}},
  \ and\ \bibinfo {author} {\bibfnamefont {J.}~\bibnamefont {Kurths}},\
  }\href@noop {} {\enquote {\bibinfo {title} {Synchronization: a universal
  concept in nonlinear science},}\ } (\bibinfo {year} {2002})\BibitemShut
  {NoStop}%
\bibitem [{\citenamefont {Acebr{\'o}n}\ \emph {et~al.}(2005)\citenamefont
  {Acebr{\'o}n}, \citenamefont {Bonilla}, \citenamefont {Vicente},
  \citenamefont {Ritort},\ and\ \citenamefont {Spigler}}]{acebron2005kuramoto}%
  \BibitemOpen
  \bibfield  {author} {\bibinfo {author} {\bibfnamefont {J.~A.}\ \bibnamefont
  {Acebr{\'o}n}}, \bibinfo {author} {\bibfnamefont {L.~L.}\ \bibnamefont
  {Bonilla}}, \bibinfo {author} {\bibfnamefont {C.~J.~P.}\ \bibnamefont
  {Vicente}}, \bibinfo {author} {\bibfnamefont {F.}~\bibnamefont {Ritort}}, \
  and\ \bibinfo {author} {\bibfnamefont {R.}~\bibnamefont {Spigler}},\
  }\href@noop {} {\bibfield  {journal} {\bibinfo  {journal} {Reviews of modern
  physics}\ }\textbf {\bibinfo {volume} {77}},\ \bibinfo {pages} {137}
  (\bibinfo {year} {2005})}\BibitemShut {NoStop}%
\bibitem [{\citenamefont {Witthaut}\ \emph {et~al.}(2022)\citenamefont
  {Witthaut}, \citenamefont {Hellmann}, \citenamefont {Kurths}, \citenamefont
  {Kettemann}, \citenamefont {Meyer-Ortmanns},\ and\ \citenamefont
  {Timme}}]{RevModPhys.94.015005}%
  \BibitemOpen
  \bibfield  {author} {\bibinfo {author} {\bibfnamefont {D.}~\bibnamefont
  {Witthaut}}, \bibinfo {author} {\bibfnamefont {F.}~\bibnamefont {Hellmann}},
  \bibinfo {author} {\bibfnamefont {J.}~\bibnamefont {Kurths}}, \bibinfo
  {author} {\bibfnamefont {S.}~\bibnamefont {Kettemann}}, \bibinfo {author}
  {\bibfnamefont {H.}~\bibnamefont {Meyer-Ortmanns}}, \ and\ \bibinfo {author}
  {\bibfnamefont {M.}~\bibnamefont {Timme}},\ }\href {\doibase
  10.1103/RevModPhys.94.015005} {\bibfield  {journal} {\bibinfo  {journal}
  {Rev. Mod. Phys.}\ }\textbf {\bibinfo {volume} {94}},\ \bibinfo {pages}
  {015005} (\bibinfo {year} {2022})}\BibitemShut {NoStop}%
\bibitem [{\citenamefont {Arenas}\ \emph {et~al.}(2008)\citenamefont {Arenas},
  \citenamefont {Díaz-Guilera}, \citenamefont {Kurths}, \citenamefont
  {Moreno},\ and\ \citenamefont {Zhou}}]{ARENAS200893}%
  \BibitemOpen
  \bibfield  {author} {\bibinfo {author} {\bibfnamefont {A.}~\bibnamefont
  {Arenas}}, \bibinfo {author} {\bibfnamefont {A.}~\bibnamefont
  {Díaz-Guilera}}, \bibinfo {author} {\bibfnamefont {J.}~\bibnamefont
  {Kurths}}, \bibinfo {author} {\bibfnamefont {Y.}~\bibnamefont {Moreno}}, \
  and\ \bibinfo {author} {\bibfnamefont {C.}~\bibnamefont {Zhou}},\ }\href
  {\doibase https://doi.org/10.1016/j.physrep.2008.09.002} {\bibfield
  {journal} {\bibinfo  {journal} {Physics Reports}\ }\textbf {\bibinfo {volume}
  {469}},\ \bibinfo {pages} {93} (\bibinfo {year} {2008})}\BibitemShut
  {NoStop}%
\bibitem [{\citenamefont {Rodrigues}\ \emph {et~al.}(2016)\citenamefont
  {Rodrigues}, \citenamefont {Peron}, \citenamefont {Ji},\ and\ \citenamefont
  {Kurths}}]{rodrigues2016kuramoto}%
  \BibitemOpen
  \bibfield  {author} {\bibinfo {author} {\bibfnamefont {F.~A.}\ \bibnamefont
  {Rodrigues}}, \bibinfo {author} {\bibfnamefont {T.~K.~D.}\ \bibnamefont
  {Peron}}, \bibinfo {author} {\bibfnamefont {P.}~\bibnamefont {Ji}}, \ and\
  \bibinfo {author} {\bibfnamefont {J.}~\bibnamefont {Kurths}},\ }\href@noop {}
  {\bibfield  {journal} {\bibinfo  {journal} {Physics Reports}\ }\textbf
  {\bibinfo {volume} {610}},\ \bibinfo {pages} {1} (\bibinfo {year}
  {2016})}\BibitemShut {NoStop}%
\bibitem [{\citenamefont {D{\"o}rfler}\ \emph {et~al.}(2013)\citenamefont
  {D{\"o}rfler}, \citenamefont {Chertkov},\ and\ \citenamefont
  {Bullo}}]{dorflerPNAS2013}%
  \BibitemOpen
  \bibfield  {author} {\bibinfo {author} {\bibfnamefont {F.}~\bibnamefont
  {D{\"o}rfler}}, \bibinfo {author} {\bibfnamefont {M.}~\bibnamefont
  {Chertkov}}, \ and\ \bibinfo {author} {\bibfnamefont {F.}~\bibnamefont
  {Bullo}},\ }\href {\doibase 10.1073/pnas.1212134110} {\bibfield  {journal}
  {\bibinfo  {journal} {Proc Natl Acad Sci U S A}\ }\textbf {\bibinfo {volume}
  {110}},\ \bibinfo {pages} {2005} (\bibinfo {year} {2013})}\BibitemShut
  {NoStop}%
\bibitem [{\citenamefont {Delabays}\ \emph
  {et~al.}(2017{\natexlab{a}})\citenamefont {Delabays}, \citenamefont
  {Coletta},\ and\ \citenamefont {Jacquod}}]{10.1063/1.4978697}%
  \BibitemOpen
  \bibfield  {author} {\bibinfo {author} {\bibfnamefont {R.}~\bibnamefont
  {Delabays}}, \bibinfo {author} {\bibfnamefont {T.}~\bibnamefont {Coletta}}, \
  and\ \bibinfo {author} {\bibfnamefont {P.}~\bibnamefont {Jacquod}},\ }\href
  {\doibase 10.1063/1.4978697} {\bibfield  {journal} {\bibinfo  {journal}
  {Journal of Mathematical Physics}\ }\textbf {\bibinfo {volume} {58}}
  (\bibinfo {year} {2017}{\natexlab{a}}),\ 10.1063/1.4978697},\ \bibinfo {note}
  {032703},\ \Eprint
  {http://arxiv.org/abs/https://pubs.aip.org/aip/jmp/article-pdf/doi/10.1063/1.4978697/15803126/032703\_1\_online.pdf}
  {https://pubs.aip.org/aip/jmp/article-pdf/doi/10.1063/1.4978697/15803126/032703\_1\_online.pdf}
  \BibitemShut {NoStop}%
\bibitem [{\citenamefont {Hindes}\ and\ \citenamefont
  {Myers}(2015)}]{10.1063/1.4927292}%
  \BibitemOpen
  \bibfield  {author} {\bibinfo {author} {\bibfnamefont {J.}~\bibnamefont
  {Hindes}}\ and\ \bibinfo {author} {\bibfnamefont {C.~R.}\ \bibnamefont
  {Myers}},\ }\href {\doibase 10.1063/1.4927292} {\bibfield  {journal}
  {\bibinfo  {journal} {Chaos: An Interdisciplinary Journal of Nonlinear
  Science}\ }\textbf {\bibinfo {volume} {25}} (\bibinfo {year} {2015}),\
  10.1063/1.4927292},\ \bibinfo {note} {073119},\ \Eprint
  {http://arxiv.org/abs/https://pubs.aip.org/aip/cha/article-pdf/doi/10.1063/1.4927292/14609862/073119\_1\_online.pdf}
  {https://pubs.aip.org/aip/cha/article-pdf/doi/10.1063/1.4927292/14609862/073119\_1\_online.pdf}
  \BibitemShut {NoStop}%
\bibitem [{\citenamefont {Wiley}\ \emph {et~al.}(2006)\citenamefont {Wiley},
  \citenamefont {Strogatz},\ and\ \citenamefont {Girvan}}]{wiley2006size}%
  \BibitemOpen
  \bibfield  {author} {\bibinfo {author} {\bibfnamefont {D.~A.}\ \bibnamefont
  {Wiley}}, \bibinfo {author} {\bibfnamefont {S.~H.}\ \bibnamefont {Strogatz}},
  \ and\ \bibinfo {author} {\bibfnamefont {M.}~\bibnamefont {Girvan}},\
  }\href@noop {} {\bibfield  {journal} {\bibinfo  {journal} {Chaos: An
  Interdisciplinary Journal of Nonlinear Science}\ }\textbf {\bibinfo {volume}
  {16}},\ \bibinfo {pages} {015103} (\bibinfo {year} {2006})}\BibitemShut
  {NoStop}%
\bibitem [{\citenamefont {Menck}\ \emph {et~al.}(2013)\citenamefont {Menck},
  \citenamefont {Heitzig}, \citenamefont {Marwan},\ and\ \citenamefont
  {Kurths}}]{menck2013basin}%
  \BibitemOpen
  \bibfield  {author} {\bibinfo {author} {\bibfnamefont {P.~J.}\ \bibnamefont
  {Menck}}, \bibinfo {author} {\bibfnamefont {J.}~\bibnamefont {Heitzig}},
  \bibinfo {author} {\bibfnamefont {N.}~\bibnamefont {Marwan}}, \ and\ \bibinfo
  {author} {\bibfnamefont {J.}~\bibnamefont {Kurths}},\ }\href@noop {}
  {\bibfield  {journal} {\bibinfo  {journal} {Nature physics}\ }\textbf
  {\bibinfo {volume} {9}},\ \bibinfo {pages} {89} (\bibinfo {year}
  {2013})}\BibitemShut {NoStop}%
\bibitem [{\citenamefont {D{\"o}rfler}\ and\ \citenamefont
  {Bullo}(2014)}]{DORFLER20141539}%
  \BibitemOpen
  \bibfield  {author} {\bibinfo {author} {\bibfnamefont {F.}~\bibnamefont
  {D{\"o}rfler}}\ and\ \bibinfo {author} {\bibfnamefont {F.}~\bibnamefont
  {Bullo}},\ }\href {\doibase https://doi.org/10.1016/j.automatica.2014.04.012}
  {\bibfield  {journal} {\bibinfo  {journal} {Automatica}\ }\textbf {\bibinfo
  {volume} {50}},\ \bibinfo {pages} {1539} (\bibinfo {year}
  {2014})}\BibitemShut {NoStop}%
\bibitem [{\citenamefont {Delabays}\ \emph
  {et~al.}(2017{\natexlab{b}})\citenamefont {Delabays}, \citenamefont {Tyloo},\
  and\ \citenamefont {Jacquod}}]{10.1063/1.4986156}%
  \BibitemOpen
  \bibfield  {author} {\bibinfo {author} {\bibfnamefont {R.}~\bibnamefont
  {Delabays}}, \bibinfo {author} {\bibfnamefont {M.}~\bibnamefont {Tyloo}}, \
  and\ \bibinfo {author} {\bibfnamefont {P.}~\bibnamefont {Jacquod}},\ }\href
  {\doibase 10.1063/1.4986156} {\bibfield  {journal} {\bibinfo  {journal}
  {Chaos: An Interdisciplinary Journal of Nonlinear Science}\ }\textbf
  {\bibinfo {volume} {27}} (\bibinfo {year} {2017}{\natexlab{b}}),\
  10.1063/1.4986156},\ \bibinfo {note} {103109},\ \Eprint
  {http://arxiv.org/abs/https://pubs.aip.org/aip/cha/article-pdf/doi/10.1063/1.4986156/13246829/103109\_1\_online.pdf}
  {https://pubs.aip.org/aip/cha/article-pdf/doi/10.1063/1.4986156/13246829/103109\_1\_online.pdf}
  \BibitemShut {NoStop}%
\bibitem [{\citenamefont {Kramers}(1940)}]{kramers1940brownian}%
  \BibitemOpen
  \bibfield  {author} {\bibinfo {author} {\bibfnamefont {H.~A.}\ \bibnamefont
  {Kramers}},\ }\href@noop {} {\bibfield  {journal} {\bibinfo  {journal}
  {Physica}\ }\textbf {\bibinfo {volume} {7}},\ \bibinfo {pages} {284}
  (\bibinfo {year} {1940})}\BibitemShut {NoStop}%
\bibitem [{\citenamefont {Dykman}(1990)}]{dykman1990large}%
  \BibitemOpen
  \bibfield  {author} {\bibinfo {author} {\bibfnamefont {M.}~\bibnamefont
  {Dykman}},\ }\href@noop {} {\bibfield  {journal} {\bibinfo  {journal}
  {Physical Review A}\ }\textbf {\bibinfo {volume} {42}},\ \bibinfo {pages}
  {2020} (\bibinfo {year} {1990})}\BibitemShut {NoStop}%
\bibitem [{\citenamefont {Hindes}\ and\ \citenamefont
  {Schwartz}(2018)}]{hindes2018rare}%
  \BibitemOpen
  \bibfield  {author} {\bibinfo {author} {\bibfnamefont {J.}~\bibnamefont
  {Hindes}}\ and\ \bibinfo {author} {\bibfnamefont {I.~B.}\ \bibnamefont
  {Schwartz}},\ }\href@noop {} {\bibfield  {journal} {\bibinfo  {journal}
  {Chaos: An Interdisciplinary Journal of Nonlinear Science}\ }\textbf
  {\bibinfo {volume} {28}},\ \bibinfo {pages} {071106} (\bibinfo {year}
  {2018})}\BibitemShut {NoStop}%
\bibitem [{\citenamefont {Hindes}\ \emph {et~al.}(2019)\citenamefont {Hindes},
  \citenamefont {Jacquod},\ and\ \citenamefont {Schwartz}}]{hindes2019network}%
  \BibitemOpen
  \bibfield  {author} {\bibinfo {author} {\bibfnamefont {J.}~\bibnamefont
  {Hindes}}, \bibinfo {author} {\bibfnamefont {P.}~\bibnamefont {Jacquod}}, \
  and\ \bibinfo {author} {\bibfnamefont {I.~B.}\ \bibnamefont {Schwartz}},\
  }\href@noop {} {\bibfield  {journal} {\bibinfo  {journal} {Physical Review
  E}\ }\textbf {\bibinfo {volume} {100}},\ \bibinfo {pages} {052314} (\bibinfo
  {year} {2019})}\BibitemShut {NoStop}%
\bibitem [{\citenamefont {Hines}\ \emph {et~al.}(2009)\citenamefont {Hines},
  \citenamefont {Balasubramaniam},\ and\ \citenamefont
  {Sanchez}}]{hines2009cascading}%
  \BibitemOpen
  \bibfield  {author} {\bibinfo {author} {\bibfnamefont {P.}~\bibnamefont
  {Hines}}, \bibinfo {author} {\bibfnamefont {K.}~\bibnamefont
  {Balasubramaniam}}, \ and\ \bibinfo {author} {\bibfnamefont {E.~C.}\
  \bibnamefont {Sanchez}},\ }\href@noop {} {\bibfield  {journal} {\bibinfo
  {journal} {Ieee Potentials}\ }\textbf {\bibinfo {volume} {28}},\ \bibinfo
  {pages} {24} (\bibinfo {year} {2009})}\BibitemShut {NoStop}%
\bibitem [{\citenamefont {Ronellenfitsch}\ \emph {et~al.}(2018)\citenamefont
  {Ronellenfitsch}, \citenamefont {Dunkel},\ and\ \citenamefont
  {Wilczek}}]{ronellenfitsch2018optimal}%
  \BibitemOpen
  \bibfield  {author} {\bibinfo {author} {\bibfnamefont {H.}~\bibnamefont
  {Ronellenfitsch}}, \bibinfo {author} {\bibfnamefont {J.}~\bibnamefont
  {Dunkel}}, \ and\ \bibinfo {author} {\bibfnamefont {M.}~\bibnamefont
  {Wilczek}},\ }\href@noop {} {\bibfield  {journal} {\bibinfo  {journal}
  {Physical Review Letters}\ }\textbf {\bibinfo {volume} {121}},\ \bibinfo
  {pages} {208301} (\bibinfo {year} {2018})}\BibitemShut {NoStop}%
\bibitem [{\citenamefont {Tyloo}\ \emph {et~al.}(2018)\citenamefont {Tyloo},
  \citenamefont {Coletta},\ and\ \citenamefont
  {Jacquod}}]{tyloo2018robustness}%
  \BibitemOpen
  \bibfield  {author} {\bibinfo {author} {\bibfnamefont {M.}~\bibnamefont
  {Tyloo}}, \bibinfo {author} {\bibfnamefont {T.}~\bibnamefont {Coletta}}, \
  and\ \bibinfo {author} {\bibfnamefont {P.}~\bibnamefont {Jacquod}},\
  }\href@noop {} {\bibfield  {journal} {\bibinfo  {journal} {Physical review
  letters}\ }\textbf {\bibinfo {volume} {120}},\ \bibinfo {pages} {084101}
  (\bibinfo {year} {2018})}\BibitemShut {NoStop}%
\bibitem [{\citenamefont {Plietzsch}\ \emph {et~al.}(2022)\citenamefont
  {Plietzsch}, \citenamefont {Auer}, \citenamefont {Kurths},\ and\
  \citenamefont {Hellmann}}]{10.1063/5.0122898}%
  \BibitemOpen
  \bibfield  {author} {\bibinfo {author} {\bibfnamefont {A.}~\bibnamefont
  {Plietzsch}}, \bibinfo {author} {\bibfnamefont {S.}~\bibnamefont {Auer}},
  \bibinfo {author} {\bibfnamefont {J.}~\bibnamefont {Kurths}}, \ and\ \bibinfo
  {author} {\bibfnamefont {F.}~\bibnamefont {Hellmann}},\ }\href {\doibase
  10.1063/5.0122898} {\bibfield  {journal} {\bibinfo  {journal} {Chaos: An
  Interdisciplinary Journal of Nonlinear Science}\ }\textbf {\bibinfo {volume}
  {32}} (\bibinfo {year} {2022}),\ 10.1063/5.0122898},\ \bibinfo {note}
  {113114},\ \Eprint
  {http://arxiv.org/abs/https://pubs.aip.org/aip/cha/article-pdf/doi/10.1063/5.0122898/16499190/113114\_1\_online.pdf}
  {https://pubs.aip.org/aip/cha/article-pdf/doi/10.1063/5.0122898/16499190/113114\_1\_online.pdf}
  \BibitemShut {NoStop}%
\bibitem [{\citenamefont {Zhang}\ \emph {et~al.}(2020)\citenamefont {Zhang},
  \citenamefont {Ma},\ and\ \citenamefont {Timme}}]{zhang2020vulnerability}%
  \BibitemOpen
  \bibfield  {author} {\bibinfo {author} {\bibfnamefont {X.}~\bibnamefont
  {Zhang}}, \bibinfo {author} {\bibfnamefont {C.}~\bibnamefont {Ma}}, \ and\
  \bibinfo {author} {\bibfnamefont {M.}~\bibnamefont {Timme}},\ }\href@noop {}
  {\bibfield  {journal} {\bibinfo  {journal} {Chaos: An interdisciplinary
  journal of nonlinear science}\ }\textbf {\bibinfo {volume} {30}},\ \bibinfo
  {pages} {063111} (\bibinfo {year} {2020})}\BibitemShut {NoStop}%
\bibitem [{\citenamefont {Tyloo}(2022)}]{tyloo2022layered}%
  \BibitemOpen
  \bibfield  {author} {\bibinfo {author} {\bibfnamefont {M.}~\bibnamefont
  {Tyloo}},\ }\href@noop {} {\bibfield  {journal} {\bibinfo  {journal} {Journal
  of Physics: Complexity}\ }\textbf {\bibinfo {volume} {3}},\ \bibinfo {pages}
  {03LT01} (\bibinfo {year} {2022})}\BibitemShut {NoStop}%
\bibitem [{\citenamefont {Martineau}\ \emph
  {et~al.}(2022{\natexlab{a}})\citenamefont {Martineau}, \citenamefont
  {Saffold}, \citenamefont {Chang},\ and\ \citenamefont
  {Ronellenfitsch}}]{Mar22}%
  \BibitemOpen
  \bibfield  {author} {\bibinfo {author} {\bibfnamefont {S.}~\bibnamefont
  {Martineau}}, \bibinfo {author} {\bibfnamefont {T.}~\bibnamefont {Saffold}},
  \bibinfo {author} {\bibfnamefont {T.~T.}\ \bibnamefont {Chang}}, \ and\
  \bibinfo {author} {\bibfnamefont {H.}~\bibnamefont {Ronellenfitsch}},\ }\href
  {\doibase 10.1103/PhysRevLett.128.098301} {\bibfield  {journal} {\bibinfo
  {journal} {Physical Review Letters}\ }\textbf {\bibinfo {volume} {128}},\
  \bibinfo {pages} {098301} (\bibinfo {year} {2022}{\natexlab{a}})}\BibitemShut
  {NoStop}%
\bibitem [{\citenamefont {DeVille}(2012)}]{deville2012transitions}%
  \BibitemOpen
  \bibfield  {author} {\bibinfo {author} {\bibfnamefont {L.}~\bibnamefont
  {DeVille}},\ }\href@noop {} {\bibfield  {journal} {\bibinfo  {journal}
  {Nonlinearity}\ }\textbf {\bibinfo {volume} {25}},\ \bibinfo {pages} {1473}
  (\bibinfo {year} {2012})}\BibitemShut {NoStop}%
\bibitem [{\citenamefont {Sch\"afer}\ \emph {et~al.}(2017)\citenamefont
  {Sch\"afer}, \citenamefont {Matthiae}, \citenamefont {Zhang}, \citenamefont
  {Rohden}, \citenamefont {Timme},\ and\ \citenamefont
  {Witthaut}}]{PhysRevE.95.060203}%
  \BibitemOpen
  \bibfield  {author} {\bibinfo {author} {\bibfnamefont {B.}~\bibnamefont
  {Sch\"afer}}, \bibinfo {author} {\bibfnamefont {M.}~\bibnamefont {Matthiae}},
  \bibinfo {author} {\bibfnamefont {X.}~\bibnamefont {Zhang}}, \bibinfo
  {author} {\bibfnamefont {M.}~\bibnamefont {Rohden}}, \bibinfo {author}
  {\bibfnamefont {M.}~\bibnamefont {Timme}}, \ and\ \bibinfo {author}
  {\bibfnamefont {D.}~\bibnamefont {Witthaut}},\ }\href {\doibase
  10.1103/PhysRevE.95.060203} {\bibfield  {journal} {\bibinfo  {journal} {Phys.
  Rev. E}\ }\textbf {\bibinfo {volume} {95}},\ \bibinfo {pages} {060203}
  (\bibinfo {year} {2017})}\BibitemShut {NoStop}%
\bibitem [{\citenamefont {Ritmeester}\ and\ \citenamefont
  {Meyer-Ortmanns}(2022)}]{Ritmeester_2022}%
  \BibitemOpen
  \bibfield  {author} {\bibinfo {author} {\bibfnamefont {T.}~\bibnamefont
  {Ritmeester}}\ and\ \bibinfo {author} {\bibfnamefont {H.}~\bibnamefont
  {Meyer-Ortmanns}},\ }\href {\doibase 10.1088/2632-072X/aca739} {\bibfield
  {journal} {\bibinfo  {journal} {Journal of Physics: Complexity}\ }\textbf
  {\bibinfo {volume} {3}},\ \bibinfo {pages} {045010} (\bibinfo {year}
  {2022})}\BibitemShut {NoStop}%
\bibitem [{\citenamefont {Tyloo}\ \emph {et~al.}(2023)\citenamefont {Tyloo},
  \citenamefont {Hindes},\ and\ \citenamefont {Jacquod}}]{tyloo2023finite}%
  \BibitemOpen
  \bibfield  {author} {\bibinfo {author} {\bibfnamefont {M.}~\bibnamefont
  {Tyloo}}, \bibinfo {author} {\bibfnamefont {J.}~\bibnamefont {Hindes}}, \
  and\ \bibinfo {author} {\bibfnamefont {P.}~\bibnamefont {Jacquod}},\
  }\href@noop {} {\bibfield  {journal} {\bibinfo  {journal} {Journal of
  Physics: Complexity}\ }\textbf {\bibinfo {volume} {4}},\ \bibinfo {pages}
  {015006} (\bibinfo {year} {2023})}\BibitemShut {NoStop}%
\bibitem [{\citenamefont {Bergen}\ and\ \citenamefont
  {Hill}(1981)}]{bergen1981structure}%
  \BibitemOpen
  \bibfield  {author} {\bibinfo {author} {\bibfnamefont {A.~R.}\ \bibnamefont
  {Bergen}}\ and\ \bibinfo {author} {\bibfnamefont {D.~J.}\ \bibnamefont
  {Hill}},\ }\href@noop {} {\bibfield  {journal} {\bibinfo  {journal} {IEEE
  transactions on power apparatus and systems}\ ,\ \bibinfo {pages} {25}}
  (\bibinfo {year} {1981})}\BibitemShut {NoStop}%
\bibitem [{\citenamefont {Martineau}\ \emph
  {et~al.}(2022{\natexlab{b}})\citenamefont {Martineau}, \citenamefont
  {Saffold}, \citenamefont {Chang},\ and\ \citenamefont
  {Ronellenfitsch}}]{PhysRevLett.128.098301}%
  \BibitemOpen
  \bibfield  {author} {\bibinfo {author} {\bibfnamefont {S.}~\bibnamefont
  {Martineau}}, \bibinfo {author} {\bibfnamefont {T.}~\bibnamefont {Saffold}},
  \bibinfo {author} {\bibfnamefont {T.~T.}\ \bibnamefont {Chang}}, \ and\
  \bibinfo {author} {\bibfnamefont {H.}~\bibnamefont {Ronellenfitsch}},\ }\href
  {\doibase 10.1103/PhysRevLett.128.098301} {\bibfield  {journal} {\bibinfo
  {journal} {Phys. Rev. Lett.}\ }\textbf {\bibinfo {volume} {128}},\ \bibinfo
  {pages} {098301} (\bibinfo {year} {2022}{\natexlab{b}})}\BibitemShut
  {NoStop}%
\bibitem [{\citenamefont {Kuramoto}(1975)}]{10.1007/BFb0013365}%
  \BibitemOpen
  \bibfield  {author} {\bibinfo {author} {\bibfnamefont {Y.}~\bibnamefont
  {Kuramoto}},\ }in\ \href@noop {} {\emph {\bibinfo {booktitle} {International
  Symposium on Mathematical Problems in Theoretical Physics}}},\ \bibinfo
  {editor} {edited by\ \bibinfo {editor} {\bibfnamefont {H.}~\bibnamefont
  {Araki}}}\ (\bibinfo  {publisher} {Springer Berlin Heidelberg},\ \bibinfo
  {address} {Berlin, Heidelberg},\ \bibinfo {year} {1975})\ pp.\ \bibinfo
  {pages} {420--422}\BibitemShut {NoStop}%
\bibitem [{Note1()}]{Note1}%
  \BibitemOpen
  \bibinfo {note} {We note that throughout this work in simulations, we draw
  the natural frequencies for each node independently from a uniform
  distribution over the interval $[-0.25,0.25]$ for simplicity.}\BibitemShut
  {Stop}%
\bibitem [{\citenamefont {Simonsen}\ \emph {et~al.}(2008)\citenamefont
  {Simonsen}, \citenamefont {Buzna}, \citenamefont {Peters}, \citenamefont
  {Bornholdt},\ and\ \citenamefont {Helbing}}]{simonsen2008transient}%
  \BibitemOpen
  \bibfield  {author} {\bibinfo {author} {\bibfnamefont {I.}~\bibnamefont
  {Simonsen}}, \bibinfo {author} {\bibfnamefont {L.}~\bibnamefont {Buzna}},
  \bibinfo {author} {\bibfnamefont {K.}~\bibnamefont {Peters}}, \bibinfo
  {author} {\bibfnamefont {S.}~\bibnamefont {Bornholdt}}, \ and\ \bibinfo
  {author} {\bibfnamefont {D.}~\bibnamefont {Helbing}},\ }\href@noop {}
  {\bibfield  {journal} {\bibinfo  {journal} {Physical review letters}\
  }\textbf {\bibinfo {volume} {100}},\ \bibinfo {pages} {218701} (\bibinfo
  {year} {2008})}\BibitemShut {NoStop}%
\bibitem [{\citenamefont {Tyloo}\ \emph {et~al.}(2019)\citenamefont {Tyloo},
  \citenamefont {Pagnier},\ and\ \citenamefont {Jacquod}}]{tyloo2019key}%
  \BibitemOpen
  \bibfield  {author} {\bibinfo {author} {\bibfnamefont {M.}~\bibnamefont
  {Tyloo}}, \bibinfo {author} {\bibfnamefont {L.}~\bibnamefont {Pagnier}}, \
  and\ \bibinfo {author} {\bibfnamefont {P.}~\bibnamefont {Jacquod}},\
  }\href@noop {} {\bibfield  {journal} {\bibinfo  {journal} {Science advances}\
  }\textbf {\bibinfo {volume} {5}},\ \bibinfo {pages} {eaaw8359} (\bibinfo
  {year} {2019})}\BibitemShut {NoStop}%
\bibitem [{\citenamefont {Mieghem}(2010)}]{mieghem_2010}%
  \BibitemOpen
  \bibfield  {author} {\bibinfo {author} {\bibfnamefont {P.~v.}\ \bibnamefont
  {Mieghem}},\ }\href {\doibase 10.1017/CBO9780511921681} {\emph {\bibinfo
  {title} {Graph Spectra for Complex Networks}}}\ (\bibinfo  {publisher}
  {Cambridge University Press},\ \bibinfo {year} {2010})\BibitemShut {NoStop}%
\bibitem [{\citenamefont {Newman}(2018)}]{newman2018networks}%
  \BibitemOpen
  \bibfield  {author} {\bibinfo {author} {\bibfnamefont {M.}~\bibnamefont
  {Newman}},\ }\href@noop {} {\emph {\bibinfo {title} {Networks}}}\ (\bibinfo
  {publisher} {Oxford university press},\ \bibinfo {year} {2018})\BibitemShut
  {NoStop}%
\bibitem [{\citenamefont {Sakaguchi}\ \emph {et~al.}(1988)\citenamefont
  {Sakaguchi}, \citenamefont {Shinomoto},\ and\ \citenamefont
  {Kuramoto}}]{sakaguchi1988mutual}%
  \BibitemOpen
  \bibfield  {author} {\bibinfo {author} {\bibfnamefont {H.}~\bibnamefont
  {Sakaguchi}}, \bibinfo {author} {\bibfnamefont {S.}~\bibnamefont
  {Shinomoto}}, \ and\ \bibinfo {author} {\bibfnamefont {Y.}~\bibnamefont
  {Kuramoto}},\ }\href@noop {} {\bibfield  {journal} {\bibinfo  {journal}
  {Progress of theoretical physics}\ }\textbf {\bibinfo {volume} {79}},\
  \bibinfo {pages} {1069} (\bibinfo {year} {1988})}\BibitemShut {NoStop}%
\bibitem [{\citenamefont {Watts}\ and\ \citenamefont
  {Strogatz}(1998)}]{watts1998collective}%
  \BibitemOpen
  \bibfield  {author} {\bibinfo {author} {\bibfnamefont {D.~J.}\ \bibnamefont
  {Watts}}\ and\ \bibinfo {author} {\bibfnamefont {S.~H.}\ \bibnamefont
  {Strogatz}},\ }\href@noop {} {\bibfield  {journal} {\bibinfo  {journal}
  {nature}\ }\textbf {\bibinfo {volume} {393}},\ \bibinfo {pages} {440}
  (\bibinfo {year} {1998})}\BibitemShut {NoStop}%
\bibitem [{\citenamefont {Freidlin}\ \emph {et~al.}(2012)\citenamefont
  {Freidlin}, \citenamefont {Sz{\"u}cs},\ and\ \citenamefont
  {Wentzell}}]{freidlin2012random}%
  \BibitemOpen
  \bibfield  {author} {\bibinfo {author} {\bibfnamefont {M.}~\bibnamefont
  {Freidlin}}, \bibinfo {author} {\bibfnamefont {J.}~\bibnamefont {Sz{\"u}cs}},
  \ and\ \bibinfo {author} {\bibfnamefont {A.}~\bibnamefont {Wentzell}},\
  }\href {http://books.google.de/books?id=p8LFMILAiMEC} {\emph {\bibinfo
  {title} {Random Perturbations of Dynamical Systems}}},\ Grundlehren der
  mathematischen Wissenschaften\ (\bibinfo  {publisher} {Springer},\ \bibinfo
  {year} {2012})\BibitemShut {NoStop}%
\bibitem [{\citenamefont {Forgoston}\ and\ \citenamefont
  {Moore}(2018)}]{doi:10.1137/17M1142028}%
  \BibitemOpen
  \bibfield  {author} {\bibinfo {author} {\bibfnamefont {E.}~\bibnamefont
  {Forgoston}}\ and\ \bibinfo {author} {\bibfnamefont {R.~O.}\ \bibnamefont
  {Moore}},\ }\href {\doibase 10.1137/17M1142028} {\bibfield  {journal}
  {\bibinfo  {journal} {SIAM Review}\ }\textbf {\bibinfo {volume} {60}},\
  \bibinfo {pages} {969} (\bibinfo {year} {2018})},\ \Eprint
  {http://arxiv.org/abs/https://doi.org/10.1137/17M1142028}
  {https://doi.org/10.1137/17M1142028} \BibitemShut {NoStop}%
\bibitem [{\citenamefont {Manik}\ \emph {et~al.}(2014)\citenamefont {Manik},
  \citenamefont {Witthaut}, \citenamefont {Sch{\"a}fer}, \citenamefont
  {Matthiae}, \citenamefont {Sorge}, \citenamefont {Rohden}, \citenamefont
  {Katifori},\ and\ \citenamefont {Timme}}]{manik2014supply}%
  \BibitemOpen
  \bibfield  {author} {\bibinfo {author} {\bibfnamefont {D.}~\bibnamefont
  {Manik}}, \bibinfo {author} {\bibfnamefont {D.}~\bibnamefont {Witthaut}},
  \bibinfo {author} {\bibfnamefont {B.}~\bibnamefont {Sch{\"a}fer}}, \bibinfo
  {author} {\bibfnamefont {M.}~\bibnamefont {Matthiae}}, \bibinfo {author}
  {\bibfnamefont {A.}~\bibnamefont {Sorge}}, \bibinfo {author} {\bibfnamefont
  {M.}~\bibnamefont {Rohden}}, \bibinfo {author} {\bibfnamefont
  {E.}~\bibnamefont {Katifori}}, \ and\ \bibinfo {author} {\bibfnamefont
  {M.}~\bibnamefont {Timme}},\ }\href@noop {} {\bibfield  {journal} {\bibinfo
  {journal} {The European Physical Journal Special Topics}\ }\textbf {\bibinfo
  {volume} {223}},\ \bibinfo {pages} {2527} (\bibinfo {year}
  {2014})}\BibitemShut {NoStop}%
\bibitem [{\citenamefont {Dekker}\ and\ \citenamefont
  {Taylor}(2013)}]{doi:10.1137/120899728}%
  \BibitemOpen
  \bibfield  {author} {\bibinfo {author} {\bibfnamefont {A.~H.}\ \bibnamefont
  {Dekker}}\ and\ \bibinfo {author} {\bibfnamefont {R.}~\bibnamefont
  {Taylor}},\ }\href {\doibase 10.1137/120899728} {\bibfield  {journal}
  {\bibinfo  {journal} {SIAM Journal on Applied Dynamical Systems}\ }\textbf
  {\bibinfo {volume} {12}},\ \bibinfo {pages} {596} (\bibinfo {year} {2013})},\
  \Eprint {http://arxiv.org/abs/https://doi.org/10.1137/120899728}
  {https://doi.org/10.1137/120899728} \BibitemShut {NoStop}%
\bibitem [{\citenamefont {Assaf}\ and\ \citenamefont
  {Meerson}(2017)}]{Assaf_2017}%
  \BibitemOpen
  \bibfield  {author} {\bibinfo {author} {\bibfnamefont {M.}~\bibnamefont
  {Assaf}}\ and\ \bibinfo {author} {\bibfnamefont {B.}~\bibnamefont
  {Meerson}},\ }\href {\doibase 10.1088/1751-8121/aa669a} {\bibfield  {journal}
  {\bibinfo  {journal} {Journal of Physics A: Mathematical and Theoretical}\
  }\textbf {\bibinfo {volume} {50}},\ \bibinfo {pages} {263001} (\bibinfo
  {year} {2017})}\BibitemShut {NoStop}%
\bibitem [{\citenamefont {Lindley}\ and\ \citenamefont
  {Schwartz}(2013)}]{lindley2013iterative}%
  \BibitemOpen
  \bibfield  {author} {\bibinfo {author} {\bibfnamefont {B.~S.}\ \bibnamefont
  {Lindley}}\ and\ \bibinfo {author} {\bibfnamefont {I.~B.}\ \bibnamefont
  {Schwartz}},\ }\href@noop {} {\bibfield  {journal} {\bibinfo  {journal}
  {Physica D: Nonlinear Phenomena}\ }\textbf {\bibinfo {volume} {255}},\
  \bibinfo {pages} {22} (\bibinfo {year} {2013})}\BibitemShut {NoStop}%
\bibitem [{\citenamefont {Ross}(2014)}]{ross2014first}%
  \BibitemOpen
  \bibfield  {author} {\bibinfo {author} {\bibfnamefont {S.~M.}\ \bibnamefont
  {Ross}},\ }\href@noop {} {\emph {\bibinfo {title} {A first course in
  probability}}}\ (\bibinfo  {publisher} {Pearson London},\ \bibinfo {year}
  {2014})\BibitemShut {NoStop}%
\bibitem [{\citenamefont {Billings}\ \emph {et~al.}(2010)\citenamefont
  {Billings}, \citenamefont {Schwartz}, \citenamefont {McCrary}, \citenamefont
  {Korotkov},\ and\ \citenamefont {Dykman}}]{billings2010switching}%
  \BibitemOpen
  \bibfield  {author} {\bibinfo {author} {\bibfnamefont {L.}~\bibnamefont
  {Billings}}, \bibinfo {author} {\bibfnamefont {I.~B.}\ \bibnamefont
  {Schwartz}}, \bibinfo {author} {\bibfnamefont {M.}~\bibnamefont {McCrary}},
  \bibinfo {author} {\bibfnamefont {A.}~\bibnamefont {Korotkov}}, \ and\
  \bibinfo {author} {\bibfnamefont {M.~I.}\ \bibnamefont {Dykman}},\
  }\href@noop {} {\bibfield  {journal} {\bibinfo  {journal} {Physical review
  letters}\ }\textbf {\bibinfo {volume} {104}},\ \bibinfo {pages} {140601}
  (\bibinfo {year} {2010})}\BibitemShut {NoStop}%
\bibitem [{\citenamefont {Dykman}(2010)}]{dykman2010poisson}%
  \BibitemOpen
  \bibfield  {author} {\bibinfo {author} {\bibfnamefont {M.}~\bibnamefont
  {Dykman}},\ }\href@noop {} {\bibfield  {journal} {\bibinfo  {journal}
  {Physical Review E}\ }\textbf {\bibinfo {volume} {81}},\ \bibinfo {pages}
  {051124} (\bibinfo {year} {2010})}\BibitemShut {NoStop}%
\bibitem [{Note2()}]{Note2}%
  \BibitemOpen
  \bibinfo {note} {For $k$-regular random networks all nodes have the same
  degree, $k$, but neighbors for each node are selected uniformly at
  random.}\BibitemShut {Stop}%
\end{thebibliography}%

\end{document}